\documentclass[12pt]{article}
\usepackage{graphicx}
\usepackage{enumerate}
\usepackage{amsmath,amssymb}
\usepackage{amsfonts}
\usepackage{cite}

\def\baselinestretch{1.3}

\newcommand{\UMNS}{U_\text{MNS}}
\DeclareMathOperator{\diag}{\text{diag}}

\renewcommand{\thefootnote}{\fnsymbol{footnote}}

\begin{document}

\def\baselinestretch{1.2}

\begin{flushright}
\footnotesize{\it
ANL-HEP-PR-07-98 \\
DESY 07-199      \\
LPT-ORSAY 07-109 \\
MAN/HEP/2007/26  \\
PITHA 07/15      }
\end{flushright}
\vspace*{-5mm}

\begin{center}
{\Large{\bf Determining Heavy Mass Parameters in  \\[2mm]
            Supersymmetric SO(10) Models}}

\vskip 15pt

{\sf
F. Deppisch$^{1,2}$\footnote[1]{E-mail:
frank.deppisch@manchester.ac.uk},
A. Freitas$^{3,4,5}$\footnote[2]{E-mail: afreitas@hep.anl.gov},
W. Porod$^6$\footnote[3]{E-mail: porod@physik.uni-wuerzburg.de}
and
P.M. Zerwas$^{1,7,8}$\footnote[4]{E-mail: zerwas@desy.de}
}
\vskip 10pt
{\it\small 
$^1$ Deutsches Elektronen-Synchrotron DESY, D-22603 Hamburg, Germany\\
$^2$ School of Physics and Astronomy,
University of Manchester, Manchester M13 9PL, UK\\
$^3$ Inst. Theor. Physik,
 Universit\"at Z\"urich, CH-8057 Z\"urich, Switzerland\\
$^4$ Enrico Fermi Institute, University of Chicago,
 Chicago, IL 60637, USA\\
$^5$ HEP Division, Argonne National Laboratory,
Argonne, IL 60439, USA\\
$^6$ Inst. Theor. Physik und Astrophysik,
 Universit\"at W\"urzburg, D-97074 W\"urzburg, Germany\\
$^7$ Inst. Theor. Physik E,
 RWTH Aachen, D-52056 Aachen, Germany\\
$^8$ Laboratoire de Physique Th{\'e}orique,
                U. Paris-Sud, F-91405 Orsay, France}
\end{center}

\begin{abstract}
Extrapolations of soft scalar mass parameters in
supersymmetric theories can be used to explore elements of the physics
scenario near the grand unification scale. We investigate the potential
of this method in the lepton sector of SO(10) which incorporates
right-handed neutrino superfields.
The method is exemplified in two models by exploring limits on the precision
that can be expected
from coherent LHC and \(e^+e^-\) collider analyses in
the reconstruction of the fundamental scalar mass parameters at the
unification scale and of the D-terms
related to the breaking of grand unification symmetries.
In addition, the mass of the third-generation right-handed neutrino
can be estimated in seesaw scenarios.
Even though the models are simplified and not intended to account for all
aspects of a final comprehensive SO(10) theory, they provide nevertheless
a valid base for identifying essential elements that can be inferred on
the fundamental high-scale theory from high-energy experiments.
\end{abstract}

\vskip 25pt

\setcounter{footnote}{0}
\renewcommand{\thefootnote}{\arabic{footnote}}

\newpage

\section{Introduction}\label{sec:introduction}

The observation of neutrino oscillations has provided experimental
proof for non-zero neutrino masses \cite{neutrino:oscillations}.
When right-handed neutrinos, not carrying any Standard Model
gauge charges, are included in the set of leptons and quarks,
the symmetry group SO(10) is naturally suggested
as the grand unification group
\cite{so10}. For theories formulated in a supersymmetric framework
to build a stable bridge between the electroweak scale and the
Planck scale, a scalar R-neutrino superfield is added to the
spectrum of the minimal supersymmetric standard model.

A natural explanation of the very light neutrino masses in
relation to the electroweak scale is offered by the seesaw
mechanism \cite{seesaw}. For right-handed Majorana neutrino masses
\(M_{\nu_{R i}}\) in a range close to the grand unification (GUT) scale, small
neutrino masses can be generated quite naturally by this
mechanism: \(m_{\nu_i}\sim m_{q_i}^2/M_{\nu_{Ri}}\), with
\(m_{q_i}\) denoting up-type quark masses. In parallel, the
R-sneutrino masses are very heavy too.

In the present analysis we will first focus on a simple model incorporating
one-step symmetry breaking
from SO(10) down to the Standard Model SM,
\begin{enumerate}
  \item[(I)]  \(\text{SO(10)} \underset{\Lambda_{\mathcal{U}}}{\to} \text{SM}\)
\end{enumerate}
with $\Lambda_{\mathcal U} \approx 2 \cdot 10^{16}$ GeV denoting
the usual GUT scale with apparent unification of the SM gauge
couplings in supersymmetric theories.
While it is not intended to account for all aspects such a model
must finally cover, we will concentrate on the analysis of a few key points
expected to be characteristic for results on the comprehensive SO(10) theory
from precision analyses at high-energy colliders.
The report expands on
earlier work in Ref.~\cite{FPZ} by including a systematic analysis
of neutrino mixing.

In a subsequent section we will add the
analysis of a specific two-step breaking chain, {\it cf.}
Ref.~\cite{Dterms},
\begin{enumerate}
  \item[(II)] \(\text{SO(10)} \underset{\Lambda_{\mathcal{O}}}{\to}
              \text{SU(5)} \underset{\Lambda_{\mathcal U}}{\to}
  \text{SM} \,,\)
\end{enumerate}
with \(\Lambda_{\mathcal{O}} > \Lambda_{\mathcal U}\) denoting the
SO(10) breaking scale and $\Lambda_{\mathcal U}$ fixed again
by the unification scale of the gauge couplings. Though this
extension will be formulated in a simple scheme, it confronts us
with problems in phenomenological analyses of
SO(10) scenarios more strongly than
the one-step scheme. Nevertheless, this hypothetical chain may
serve as an interesting example for elucidating to what extent
the high complexity encountered in SO(10) scenarios can be
controlled eventually.

\begin{figure}[t]
\centering
\includegraphics[clip,width=0.45\textwidth]{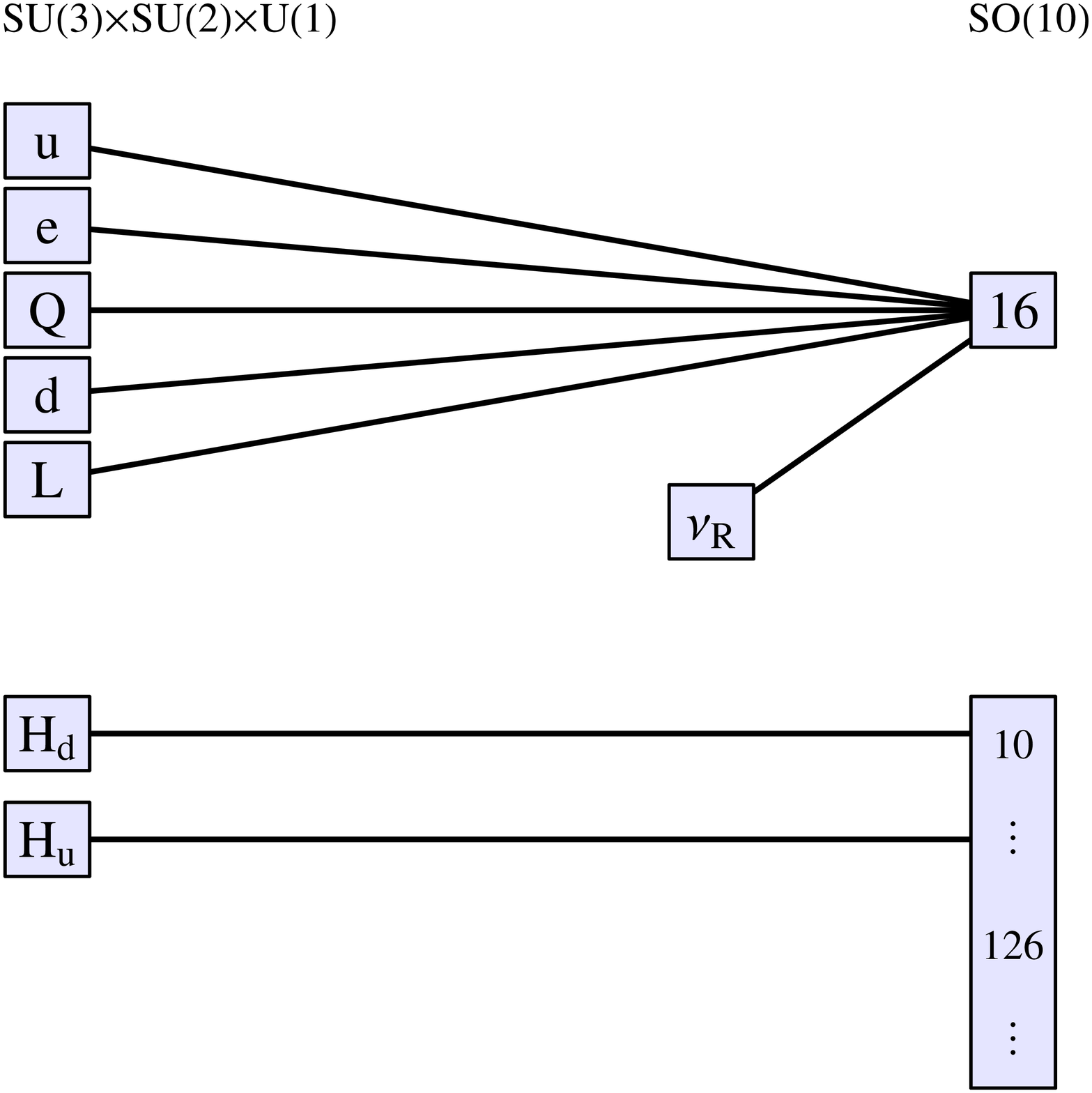} \hspace*{10mm}
\includegraphics[clip,width=0.45\textwidth]{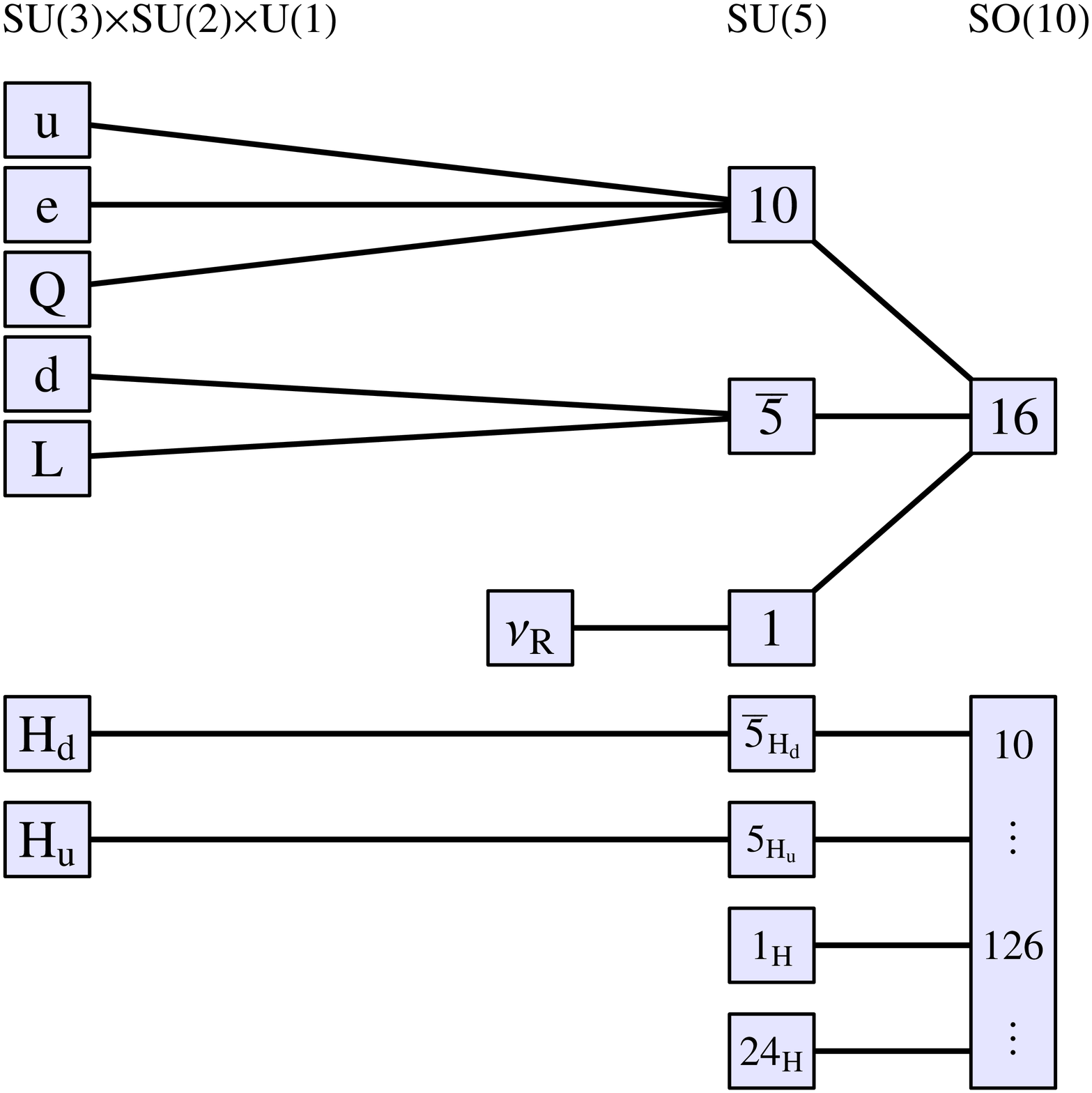}
     \caption{\it   Potential decompositions
                    of SO(10) representations, {\rm Left:} one-step breaking
                    $SO(10) \to SU(3) \times SU(2) \times U(1)$;
                    {\rm Right:} two-step breaking
                    $SO(10) \to SU(5) \to SU(3) \times SU(2) \times
                    U(1)$. The dots denote Higgs fields
                    for generating masses of matter fields and breaking the SO(10)        
                    gauge symmetry in addition to those which are indicated explicitly 
                    in standard notation.}
     \label{fig:BreakingDiagrams}
\end{figure}

{\bf (I)} In the first, one-step scenario the scalar soft SUSY
breaking sector, {\it cf.} Fig.~\ref{fig:BreakingDiagrams}(left),
is parametrized by the gravity induced mass parameters
for the matter superfields and for the
Higgs superfields at the unification scale $\Lambda_{\mathcal{U}}$.
Two Higgs fields
generate masses separately for up- and down-type
fields at the electroweak scale.
In general they evolve into a superposition of iso-doublet components
in an ensemble of Higgs fields at the grand unification scale \cite{A1,G1}.
An extended ensemble of Higgs fields in large representations is needed
in renormalizable theories to account for the mass matrices of all three
generations, {\it cf.} Refs.~\cite{A1,A2,Aulakh:2006hs,G1} [Generally, though, 
this does not solve
the fine-tuning problem of the light Higgs fields]. This expansion 
however is not necessarily required for the approximate description of the third
generation which dominates our analysis as a result of the large Yukawa
couplings, and we will assume the standard Higgs-10 field to be dominant.
Alternatively the Higgs sector may be supplemented
by Planck-scale suppressed operators to account for mass spectra and
mixing phenomena \cite{Wies,Nath}. In the present context the mechanism
for generating the masses of the right-handed Majorana neutrinos
needs not be specified in detail as long as the scale is close to the
grand unification scale. A popular choice is a Higgs-126 field
with small couplings to the $\nu_R$ fields \cite{Babu}.

The scalar mass parameters
are assumed universal:
\begin{equation}\label{eq:ScalarUniversality}
m_{16} = m_{10} = ... =M_0 \,.
\end{equation}
The GUT scale \(\Lambda_{\mathcal U}\) is defined technically by
the minimum distance of the gauge couplings within the triangle
built by the running couplings near the unification point.  At
\(\Lambda_{\mathcal U}\) the soft scalar mass parameters are
shifted, non-universally, by D-terms associated with the SO(10)
breaking to the lower-rank SM group \cite{DreesDT,Kolda:1996iw}.
Starting at \(\Lambda_{\mathcal U}\), the mass parameters evolve,
following the renormalization group (RG) \cite{JJK}, down to the
electroweak scale. They define the Lagrangian parameters at the
supersymmetry scale \(\tilde M\), chosen at 1~TeV according to the
SPA convention \cite{spa}. The observed masses of sleptons and
squarks, charginos/neutralinos and Higgs bosons can be expressed
in terms of these Lagrangian parameters
\cite{Pierce:1996zz,MartinSP}. Once the masses are measured, the
RG evolution from the Terascale upwards will allow us to
reconstruct the physics scenario at the GUT scale
\cite{FPZ,GUTreconstruction,GUTreconstruction2}.

For the mass parameters of matter fields in the first two
generations and the Higgs field $H_d$ the RG flow from the
Terascale to the GUT scale is, effectively, not interrupted by
any intermediate thresholds. The first two generations and the Higgs
$H_d$ can therefore be exploited to extract the scalar GUT mass parameters
\(m_{16}\) and \(m_{10}\), and the D-term. By contrast, the
running of the mass parameters in the third generation and of the
second $H_u$ Higgs mass parameter are
affected by Yukawa interactions at the intermediate seesaw scale
\cite{Baer:2000hx,GUTreconstruction2,FPZ}. The resolution of
\(\nu_R\)-higgsino and ${\tilde{\nu}}_R$-Higgs loops gives rise to
kinks in the evolution of the iso-doublet scalar L-mass parameter
and the $H_u$ Higgs mass parameter. Linking the measured
\(\tilde\tau\) and $\tilde{\nu}_\tau$ masses to the GUT parameter
$m_{16}$, assumed universal, determines the position of the kink
and thus allows us to measure the R-neutrino mass, i.e. the seesaw
scale of the third generation.

{\bf (II)} In the second scenario the GUT symmetry is broken down
to the Standard Model in two consecutive steps, {\it cf.}
Fig.~\ref{fig:BreakingDiagrams}(right). At the scale
$\Lambda_{\mathcal{O}} > \Lambda_{\mathcal U}$ the SO(10) symmetry
breaks to SU(5). Apart from the D-terms, the scalar mass parameter
\(m_{16}\) splits at $\Lambda_\mathcal{O}$ to three separate
parameters \(m_{10}\), \(m_{\bar{5}}\) and \(m_1\) according to
the decomposition of the matter multiplet \(16=10+\bar 5+1\). At
the scale \(\Lambda_{\mathcal U}\) the SU(5) gauge symmetry is broken
to the SM symmetry by a $\{24\}$ Higgs field, for example.
At \(\Lambda_{\mathcal U}\) the scalar
SU(5) mass parameters finally split to the MSSM parameters
evolving down to the Terascale. To simplify the analysis, we
assume all the Higgs mass parameters
to be degenerate with \(m_{16}\) at \(\Lambda_{\mathcal{O}}\).
Such an assumption can be motivated by string theories, in which
the scales $\Lambda_{str}$ and $\Lambda_{\mathcal{O}}$ are
identified \cite{Dienes:1996du}.
The evolution between $\Lambda_{\mathcal U}$ and the Terascale is
driven by {\it a priori} experimental information, modulo the
effect of the heavy R-neutrino mass parameter of the third
generation. By contrast, the evolution from \(\Lambda_{\mathcal{O}}
\to \Lambda_{\mathcal U}\) depends on the field content of the
high-scale SU(5) theory.
To determine, {\it a posteriori},
such elements of the physics
scenario at the high scales by
means of renormalization group techniques
is the prime target of extrapolations
from the experimentally accessible Terascale.

The analyses assume, implicitly, that a number of fundamental
problems \cite{Raby} are solved without interfering strongly with
the present parametric analysis, {\it i.e.} mechanisms leading to
doublet-triplet splitting and suppressing proton decay. Solutions,
partial in general, are approached by means of extended Higgs
sectors. Extended Higgs systems have also been shown to accommodate 
the fermion mass patterns in renormalizable theories, {\it cf.} 
Refs.~\cite{A1,A2,Aulakh:2006hs,G1}. Alternatively the Higgs system is supplemented 
by higher-dimensional, Planck-scale suppressed operators, 
{\it cf.} Refs.~\cite{Wies,Nath}. In addition, if the survival principle 
is violated \cite{surv}, potential contributions of intermediate states 
have to be taken into account. When these mechanisms are
specified, also parametrically, it is straightforward to
transcribe the results we will derive to any such system provided
the additional interactions, giving rise to threshold effects on
the observables, can be treated {\it ad hoc} as a perturbation. If
not, the evolution of the scalar masses must be reformulated by
including the new degrees of freedom from the beginning. Once such
a system is defined properly, the analysis can be performed
strictly in parallel to the procedure elaborated in the present
paper.

Though only a limited set of elements in the
high-scale scenario can be focused on, the potential of such
analyses in exploring the high-scale scenario can nevertheless be
elucidated.
Refinements and modifications can be incorporated if
the theoretical frame for solving the problems mentioned above,
is specified {\it in toto}. In this context, the results presented
can serve as a crucial intermediate step
for the analysis of a comprehensive SO(10) theory.

Given the multitude of potential physics scenarios
near the Planck scale, these experimental projections --
supplemented by other observations in the neutrino sector, proton
decay, lepton flavor violating processes, and cosmological
observations -- will be a valuable ingredient for reconstructing
the fundamental high-scale theory.

\section{One-Step SO(10) \(\to\) SM Breaking}\label{sec:onestep} 

In the SO(10) model which we will analyze, the matter superfields
of the three generations belong to 16-dimensional representations
of SO(10) and the standard Higgs superfield coupling
to the up-type matter fields of the third generation is embedded
in a Higgs-10 field at the unification scale. No reference is needed
to the Yukawa couplings of the first two generations.
The mass of the heavy R-neutrino superfields may be
generated by a 126-dimensional Higgs field though the detailed
mechanism actually needs not be specified as long as the
characteristic scale of the mechanism is connatural to the
unification scale.

In this set-up the Yukawa couplings in the neutrino sector coincide
with the Yukawa couplings in the up-type quark matrix. Assuming
the normal hierarchy for the light neutrino masses and approximate
tri-bimaximal mixing, as suggested experimentally, the texture of
the heavy Majorana mass matrix is predicted within the seesaw
mechanism. In this framework the evolution of the soft scalar slepton
mass parameters can be predicted from the unification scale down
to the electroweak scale. Since the right-handed neutrino fields
are neutral under the SM gauge group, they merely affect the evolution by Yukawa
interactions which are sufficiently large only in the L-sector of the
third generation as well as the $H_u$ Higgs
sector.

The scalar mass parameters $m_{16}$ and the Higgs parameters
\(m_{10}\) and \(m_{10'}\) at the unification scale will be
assumed universal, {\it cf.} Eq.~(\ref{eq:ScalarUniversality}).
However, the breaking of the rank-5 SO(10) symmetry group to the
lower rank-4 SM group generates GUT D-terms $D_{\mathcal{U}}$ such that the
boundary conditions at the GUT scale finally read
\cite{Cheng:1994bi,Dterms} for the matter fields,
\begin{eqnarray}
  m_L^2     &=& M_0^2   - 3D_{\mathcal{U}}  \nonumber\\
  m_E^2     &=& M_0^2   +  D_{\mathcal{U}}  \nonumber\\
  m_R^2     &=& M_0^2   + 5D_{\mathcal{U}}  \,,
\end{eqnarray}
and for the Higgs fields,
\begin{eqnarray}
  m_{H_d}^2 &=& M_0^2 + 2D_{\mathcal{U}}             \nonumber\\
  m_{H_u}^2 &=& M_0^2 - 2D_{\mathcal{U}}    \,.
\label{eq:Hu-Dterm}
\end{eqnarray}
The L-isodoublet, the charged R-isosinglet, the neutral
R-isosinglet and the two Higgs scalar mass parameters, are denoted
by \(m_L, m_E, m_R, m_{H_{d,u}}\), respectively. It can be shown
on general grounds that the D-term is of the order of the soft
SUSY breaking masses of the fields responsible for the spontaneous
breaking times the gauge coupling squared \cite{Kolda:1996iw},
\begin{equation}\label{eq:Dterm}
D_{\mathcal{U}}\sim g^2_{SO(10)} \, \mathcal{O}(M_0^2) \,,
\end{equation}
while the detailed form depends on the specific SO(10) breaking
mechanism. Not specifying the structure of this component of the Higgs
sector, we will treat $D_{\mathcal{U}}$ as a free parameter.  
While the coefficients of the D-terms for the matter parameters are
fixed uniquely, they are in general model-dependent for the Higgs
fields \cite{Aulakh:2006hs}, which however play a minor role in our
analysis.  The evolution of the scalar masses \(m^2_{L,E}\) from the
unification scale down to the Terascale scale is determined by
lepton-gaugino and slepton-gauge loops, complemented by
R-neutrino-higgsino loops {\it etc.} in the third generation. \\[-2mm]

{\it a) The neutrino sector:} \\[3mm]
Neglecting minor higher-order effects in the
calculation of the Majorana neutrino mass matrix, it follows from
the Higgs-10 SO(10) relation
\begin{equation}\label{eqn:YukawaUnification}
    Y_\nu = Y_u
\end{equation}
between the neutrino and up-type quark Yukawa matrices that
\begin{equation}\label{eqn:YnuApproximation}
    Y_\nu \approx \diag(m_u,m_c,m_t)/v_u
\end{equation}
holds approximately for the neutrino Yukawa matrix; $v_u =
v\sin\beta$, with $v$ and $\tan\beta$ being the familiar vacuum
and mixing parameters in the Higgs sector. The quark masses are
defined at the scale $\Lambda_{\mathcal U}$.
The identity Eq.~(\ref{eqn:YukawaUnification})
is assumed for the third generation
while the more complex mass pattern in the first two generations
is essentially ineffective in the numerical analysis.
Quark mixing and RG
running effects in the neutrino sector are neglected in the
analytical approach but properly taken into account in the
numerical analysis.

The effective mass matrix of the light
neutrinos is constrained by the results of the neutrino
oscillation experiments:
\begin{equation}\label{eqn:mnu}
    m_\nu=
    \UMNS^* \cdot
    \diag(m_{\nu_1},m_{\nu_2},m_{\nu_3}) \cdot
    \UMNS^\dagger.
\end{equation}
We will assume the normal hierarchy for the light neutrino
masses $m_{\nu_i}$, and for the MNS mixing matrix the
tri-bimaximal form
\begin{equation}\label{eqn:UMNS}
    \UMNS =
    \begin{pmatrix}
       \sqrt{2/3} & \sqrt{1/3} &           0 \\
      -\sqrt{1/6} & \sqrt{1/3} &  \sqrt{1/2} \\
      -\sqrt{1/6} & \sqrt{1/3} & -\sqrt{1/2} \\
    \end{pmatrix}.
\end{equation}
From the seesaw relation
\begin{equation}\label{eqn:SeesawRelation}
    M_{{\nu}_R} = Y_\nu m_\nu^{-1} Y_\nu^T \cdot v^2_u,
\end{equation}
the heavy Majorana R-neutrino mass matrix $M_{{\nu}_R}$ can
finally be derived as
\begin{equation}\label{eqn:MApproximation}
    M_{{\nu}_R} \approx
    \diag(m_u,m_c,m_t) m_\nu^{-1}\diag(m_u,m_c,m_t).
\end{equation}

Solving Eq.~(\ref{eqn:MApproximation}) for the eigenvalues
$M_{\nu_{Ri}}$ $(i=1,2,3)$, the heavy Majorana masses are
determined by the up-quark masses $m_{u,c,t}$ at the GUT scale and
the light neutrino masses \(m_{\nu_1}, m_{\nu_2}, m_{\nu_3}\). For
normal hierarchy, \(m_{\nu_3}\) and \(m_{\nu_2}\) are given by
\(m_{\nu_3}=\sqrt{m_{\nu_1}^2+\Delta m_{13}^2}\) and
\(m_{\nu_2}\approx \sqrt{m_{\nu_1}^2 + \Delta m_{12}^2}\), with
the mass squared differences \(\Delta m_{13}^2\approx 2\cdot
10^{-3}\,\text{eV}^2\) and \(\Delta m_{12}^2\approx 8\cdot
10^{-5}\,\text{eV}^2\), measured in neutrino oscillation
experiments. The $M_{{\nu}_{R i}}$ spectrum is then predicted as a
function of the masses of the light neutrinos.
\begin{figure}[t]
\centering
\includegraphics[clip,width=0.7\textwidth]{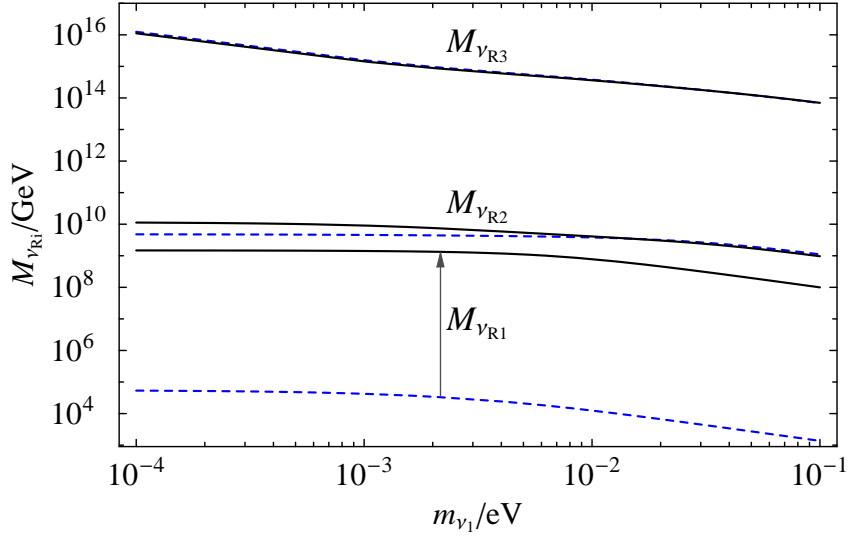}
     \caption{\it Masses of right-handed neutrinos $M_{\nu_{Ri}}$
     as functions of the lightest neutrino mass \(m_{\nu_1}\).
     The dashed (blue) lines assume perfect Yukawa unification,
     Eq.~(\ref{eqn:YukawaUnification}).
     The solid (black) lines indicate the shifts of the $\nu_R$ masses in the first and
     and second generation if the Yukawa identity Eq.~(\ref{eqn:YukawaUnification})
     is modified {\it ad-hoc} by a term $\kappa/v$ with $\kappa = 100$~{\rm MeV}. }
     \label{fig:Mivsm1}
\end{figure}
Quite generally, the solution for the eigenvalues can be
approximated to a high level of accuracy by the relations [see
also Ref.~\cite{Smir}]:
\begin{eqnarray}\label{eqn:EnhancedApproximation}
        M_{{\nu}_{R 1}} &\approx&
             \frac{m_{\nu_1}+2m_{\nu_2}+3m_{\nu_3}}{3m_{\nu_1} m_{\nu_2}+ 2 m_{\nu_1} m_{\nu_3}+ m_{\nu_2} m_{\nu_3}}
             \, m_u^2  \nonumber \\
        M_{{\nu}_{R 2}} &\approx&
             \frac{4\,m_{\nu_1}+2\,m_{\nu_2}+0\,m_{\nu_3}}{3m_{\nu_1} m_{\nu_2} + 2m_{\nu_1} m_{\nu_3}
             + m_{\nu_2} m_{\nu_3}}\, m_c^2     \\
        M_{{\nu}_{R 3}} &\approx&
             \frac{3m_{\nu_1} m_{\nu_2} + 2m_{\nu_1} m_{\nu_3} + m_{\nu_2} m_{\nu_3}}
             {6m_{\nu_1} m_{\nu_2} m_{\nu_3}}\,
             m_t^2  \,. \nonumber
\end{eqnarray}
Thus the mass spectrum of the R-neutrinos is strongly ordered in SO(10) with
minimal Higgs content, \(M_{\nu_{R3}}:M_{\nu_{R2}}:M_{\nu_{R1}} \sim
m_t^2:m_c^2:m_u^2\).

The numerical evaluation, including refinements like RG running
effects, is displayed in Fig.~\ref{fig:Mivsm1} for a wide range of
$m_{\nu_1}$ values. The analytical approximation,
Eq.~(\ref{eqn:EnhancedApproximation}), is very accurate across the
entire range, from small $m_{\nu_1}$ with strong ordering of the
hierarchical light neutrino masses, $m_{\nu_1} \ll m_{\nu_2} \ll
m_{\nu_3}$, up to nearly degenerate light neutrino masses,
$m_{\nu_i} \to m_{\nu_1}$. The heavy R-neutrino masses read
\begin{equation}
    M_{\nu_{Ri}} =
    \begin{cases}
    \,(      3m^2_u/m_{\nu_2}, \,      2m^2_c/m_{\nu_3}, \,\frac{1}{6}m^2_t / m_{\nu_1}), & m_{\nu_1}\ll \sqrt{\Delta m_{12}^2}\\
    \,(\,\,\, m^2_u/m_{\nu_1}, \,\,\,\, m^2_c/m_{\nu_1}, \,\,\,\,     m^2_t / m_{\nu_1}), & m_{\nu_1}\gg \sqrt{\Delta m_{13}^2}
    \end{cases}
\end{equation}
in these two limits.

It should be noted that the prediction for the third-generation
R-neutrino mass $M_{\nu_{R3}}$ is quite robust, contrary to the
second and the first generation in particular. Modifying the
relation between the neutrino and up-type quark Yukawa couplings,
Eqs.~(\ref{eqn:YukawaUnification}), {\it ad-hoc} by a small
additional term, $Y_\nu=Y_u+\kappa/v$, with $\kappa \sim $
a few hundred MeV associated potentially with a more complex Higgs
scenario, Planck-scale suppressed contributions or
non-perturbative effects at small mass scales, the first
generation R-neutrino mass is lifted to $\sim 10^9$ GeV. This just
illustrates that a small modification of the $\nu$-$up\;quark$
Yukawa identity is sufficient to reconcile the mass estimate with
limits suggested within leptogenesis scenarios
for the matter-antimatter asymmetry in the Universe \cite{Leptogen}.

The small mixing of the right-handed neutrinos,
\begin{equation}
    U_R \approx
    \begin{pmatrix}
                                   1 &  a &  \mathcal{O}(\frac{m_u}{m_t}) \\
                                  -a &  1 &                             b \\
        \mathcal{O}(\frac{m_u}{m_t}) & -b &                             1 \\
    \end{pmatrix} \;
    \quad
    \begin{aligned}
      a &= \frac{-2(m_{\nu_1}-m_{\nu_2})m_{\nu_3}}{3m_{\nu_1}m_{\nu_2}
           +\,\,\,2m_{\nu_1} m_{\nu_3}+m_{\nu_2} m_{\nu_3}} \frac{m_u}{m_c}
         = \mathcal{O}\left(\frac{m_u}{m_c}\right)                                           \\
      b &= \frac{3m_{\nu_1} m_{\nu_2}-2m_{\nu_1} m_{\nu_3}-2m_{\nu_2} m_{\nu_3}}
           {(3m_{\nu_1}+m_{\nu_3})m_{\nu_2}}\frac{m_c}{m_t}
         = \mathcal{O}\left(\frac{m_c}{m_t}\right),
    \end{aligned}
\end{equation}
hardly affects the analysis, since it enters only indirectly into
the RG evolution of the slepton mass parameters when the
right-handed
neutrinos are decoupled at their own mass scale.

The Yukawa mass matrix squared, which determines the connection of
the slepton masses in the third generation at low and high scales,
is dominated by the 33 element,
\begin{equation}\label{eqn:YY33Element}
    \left(Y_\nu^\dagger Y_\nu\right)_{33} \approx m_t^2(\Lambda_{\mathcal U})/v^2_u \approx 0.3 \,,
\end{equation}
while the other elements are suppressed to a level of \(10^{-2}\)
down to \(10^{-5}\). \\[-2mm]

{\it b) Scalar mass parameters:} \\[3mm]
The slepton systems of the \emph{first two generations} depend, to
leading order, on four parameters: the two fundamental scalar
and gaugino mass parameters, the D-term and the unification scale. They can be
uniquely determined from the unification scale of the gauge
couplings, the L- and R-slepton masses and the masses of the
charginos and neutralinos. The complementary analysis of squarks
provides an independent cross-check of the underlying picture.

Analytic relations, for the sake of clarity, are presented only
to leading logarithmic order
while the subsequent numerical analyses include the
non-logarithmic higher order contributions. To leading order,
the solutions of
the RG equations, the masses of the scalar selectrons and the
L-type $e$-sneutrino, can be expressed in terms of the high scale
parameter $m_{16} = M_0$, the universal gaugino mass parameter
\(M_{1/2}\) and the GUT and electroweak D-terms, $D_{\mathcal{U}}$ and
\(D_{EW}=M_Z^2/2 \, \cos2\beta\), respectively:
\begin{eqnarray}
  m_{\tilde e_R}^2     &=& M_0^2  +  D_{\mathcal{U}} + \alpha_R M_{1/2}^2 - \tfrac{6}{5}S' - 2s_W^2   D_{EW} \nonumber  \\
  m_{\tilde e_L}^2     &=& M_0^2  - 3D_{\mathcal{U}} + \alpha_L M_{1/2}^2 + \tfrac{3}{5}S' - c_{2W}   D_{EW} \nonumber \\
  m_{\tilde\nu_{eL}}^2 &=& M_0^2  - 3D_{\mathcal{U}} + \alpha_L M_{1/2}^2 + \tfrac{3}{5}S' +          D_{EW} \,,
  \label{eq:RGfirstgen}
\end{eqnarray}
[as usual, $s^2_W = \sin^2\theta_W$ {\it etc}.].
This set of relations is valid under the assumption of small
threshold corrections at the grand unification scale. Though the observables
are different, the assumption is backed qualitatively nevertheless
by the strongly suggested unification of the gauge couplings 
\cite{GUTreconstruction2} which does not require sizeable contributions 
from thresholds or intermediate-scale degrees of freedom. 
It has been explicitly shown that the threshold effects 
due to the large SO(10) representations are small 
and do not destroy gauge coupling unification \cite{Aulakh:2007jm}.

The coefficients
\(\alpha_L\) and \(\alpha_R\) are given by the
gaugino/gauge boson loops in the RG evolution from the global
supersymmetry scale $\tilde{M}$ \cite{spa} to the unification
scale $\Lambda_{\mathcal U}$,
\begin{eqnarray}
  \alpha_L &=& \tfrac{3}{10} f_1 + \tfrac{3}{2} f_2     \nonumber \\
  \alpha_R &=&  \tfrac{6}{5} f_1  \\
   f_i &=& \frac{1}{b_i} \left(
1 - {\left[1 + \frac{\alpha_{\mathcal U}}{4 \pi} b_i \log
\frac{\Lambda_{\mathcal U}^2}{{\tilde{M}}^2}\right]^{-2}} \right)
\nonumber \,\, \,\, {\rm with} \,\, \,\,(b_1,b_2) = \textstyle
(\frac{33}{5},1) \, ,
\end{eqnarray}
and the numerical evaluation finally yields \(\alpha_R \approx 0.15\)
and \(\alpha_L \approx 0.5\) for $\tilde{M} = 1$ TeV.
The universal gaugino mass parameter
$M_{1/2}$ can be pre-determined in the chargino/neutralino sector.
The non-universal initial conditions in the evolution due to the
D-terms generate the small generation-indepen\-dent corrections
\begin{equation}
S' = - 4 D_{\mathcal{U}} \,
\frac{\alpha_1(\tilde{M})}{\alpha_1(\Lambda_{\mathcal U})} \,,
\end{equation}
{\it cf.} Ref.~\cite{Cheng:1994bi}.

The Higgs parameter $m_{10} = M_0$
can be calculated analogously,
\begin{equation}
m^2_{H_d} = M_0^2 + 2 D_{\mathcal{U}} + \alpha_L M_{1/2}^2
+\tfrac{3}{5}S' - \Delta_\tau - 3 \Delta_b \, .
\label{eq:Hd-Dterm}
\end{equation}
$\Delta_\tau$ and $\Delta_b$ describe contributions involving
$\tau$ and $b$ loops, respectively, with correspondingly large
Yukawa couplings, see Eqs.~(\ref{eq:DeltaTau}) and (\ref{eq:DeltaBottom}) below.

The scalar masses of the \emph{third generation} receive
additional contributions from $\nu_{R\tau}$-higgsino loops {\it
etc.}, coupled by Yukawa interactions with the L and R fields. The
masses of the third generation are shifted relative to the masses
of the first two generations by two terms
\cite{GUTreconstruction2,FPZ}:
\begin{eqnarray}
  m_{\tilde \tau_R}^2 - m_{\tilde e_R}^2    &=& m_\tau^2 - 2\Delta_\tau                    \nonumber \\
  m_{\tilde \tau_L}^2 -  m_{\tilde e_L}^2  &=& m_\tau^2 -  \Delta_\tau - \Delta_{\nu_\tau} \nonumber  \\
  m_{\tilde\nu_{\tau L}}^2 -  m_{\tilde\nu_{eL}}^2 &=& -  \Delta_\tau - \Delta_{\nu_\tau}    \,.
  \label{eq:RGthirdgen}
\end{eqnarray} 
[It should be emphasized that such differences, which are crucial
for controlling the R-neutrino mass of the third generation,
are generally more immune to corrections than the individual observables.]

The Higgs parameter $m^2_{H_u}$ is given analogously, in the
universality class $m_{H_u} = M_0$, by
\begin{equation}
      m^2_{H_u} = M_0^2 - 2 D_{\mathcal{U}} -\tfrac{3}{5}S' + \alpha_L M_{1/2}^2
      - \Delta_{\nu_\tau} - 3 \Delta_t \,.
\end{equation}

The shifts $\Delta_\tau$ and $\Delta_{\nu_\tau}$, generated by
loops \cite{H} involving charged lepton and neutrino superfields,
respectively, are predicted by the renormalization group in the
SO(10) scenario,
\begin{eqnarray}
  \Delta_\tau       &\approx&
    \frac{m_\tau^2(\Lambda_{\mathcal U})}{8 \pi^2
        v_d^2}\left(3M_0^2+A_0^2\right)\log\frac{\Lambda_{\mathcal U}^2}{\tilde{M}^2} \label{eq:DeltaTau}\\
  \Delta_{\nu_\tau} &\approx&
        \frac{m_t^2(\Lambda_{\mathcal U})}{8 \pi^2
        v_u^2}\left(3M_0^2+A_0^2\right)\log\frac{\Lambda_{\mathcal U}^2}{M_{\nu_{R3}}^2} \, ,
  \label{eq:DeltaNuTau}
\end{eqnarray}
including the [universal] trilinear coupling $A_0$, while
\begin{eqnarray}
  \Delta_b &\approx& \frac{m^2_b(\Lambda_{\mathcal U})}{8\pi^2 v^2_d}\left(3M_0^2+A_0^2\right)\log\frac{\Lambda^2_{\mathcal U}}{\tilde{M}^2}\label{eq:DeltaBottom}\\
  \Delta_t &\approx& \frac{m_t^2(\Lambda_{\mathcal U})}{8\pi^2 v^2_u}\left(3M_0^2+A_0^2\right)\log\frac{\Lambda^2_{\mathcal U}}{\tilde{M}^2}\label{eq:DeltaTop}\,
\end{eqnarray}
account similarly for the running from the Tera- to the GUT scale
induced by $b,t$ loops. The parameter \(m_t(\Lambda_{\mathcal
U})=m_t(v)(\Lambda_{\mathcal U}/v)^\gamma\) is the running top
quark mass with, approximately,
\(\gamma=(6y_t^2-16/3g_s^2)/(16\pi^2)\), the top-quark Yukawa
coupling $y_t$ and the QCD coupling $g_s$ evaluated at the
electroweak scale $v$; $v_{d(u)} = v \cos\beta \, (v \sin\beta)$.

While the slepton masses,
specifically the sum $m^2_{{\tilde{\tau}}_L} + m^2_{{\tilde{\tau}}_R} =
m^2_{{\tilde{\tau}}_1} + m^2_{{\tilde{\tau}}_2}$ of the third generation,
are experimentally accessible directly, the
Higgs mass parameters,
\begin{equation}
m^2_{H_{d(u)}} = M_A^2 \sin^2\beta \cos^2\beta - |\mu|^2
                                \mp \tfrac{1}{2} M_Z^2 \cos 2 \beta
                                \label{eq:higgsrel}
\end{equation}
can be determined from the pseudo-scalar Higgs-boson mass $M_A$
and the higgsino mass parameter $\mu$. The effect of the
Higgs sector on the GUT parameters is small so that potential
modifications of D-terms
in Eqs.~(\ref{eq:Hu-Dterm},\ref{eq:Hd-Dterm}) 
do not have a significant impact. \\[-2mm]

{\it c) Numerical results:}\\[3mm]
Anticipating high-precision measurements at future colliders,
such an SO(10) scenario can be investigated in central facets. As
a concrete example, we study the following LR-extended
scenario with MSSM parameters close to SPS1a/a$'$ \cite{SPS,spa}:
\begin{equation}
\begin{array}{rclrcl}
   M_0 &=& 90 {\rm{\ GeV}} \hspace{1.5cm}
   & \tan\beta &=& 10  \\
   M_{1/2}       &=& 250 {\rm{\ GeV}} & {\rm sgn}(\mu)     &=& +         \\
   A_0           &=& -640 {\rm{\ GeV}}
\end{array}
\end{equation}
and
\vspace*{-5mm}
\begin{eqnarray}
   D_{\mathcal{U}}          &=& -0.9 \cdot 10^3 {\rm{\ GeV}}^2       \nonumber \\
   M_{\nu_{R3}} &=& 7.2 \cdot 10^{14} {\rm{\ GeV}}       \, .
\end{eqnarray}
The low-energy and cosmological data of this scenario are
compatible with observations: QCD coupling $\alpha_s(M_Z)=0.119$,
electroweak mixing parameter
${\sin^2\theta}^{lept}_{eff}=0.23140$, radiative $b$ decay BR$(b
\to s \gamma) = 3.32 \cdot 10^{-4}$, deviation of the muon
anomalous magnetic moment from SM value $\Delta a_\mu = \Delta
(g_\mu - 2)/2 = 15.3 \cdot 10^{-10}$, and cold-dark-matter density
$\Omega h_0^2 = 0.110 $. [Introducing just one Higgs-10
field, giving rise to the unification of the $t-b-\tau$ Yukawa
couplings, would require a larger value of $\tan\beta$. Extended
Higgs sectors at the GUT scale avoiding this constraint can be
designed without modifying the present analysis.]

The estimated experimental precision for the
measurements of slepton, sneutrino and gaugino masses at LHC and
ILC can be extrapolated from the results of
Refs.~\cite{a,FPZ,c,c1,spa,d,Bechtle:2004pc},
which are based on detailed simulations.
\begin{table}
  \centering
  \begin{tabular}{|l|l|l||l|l@{}l|l|}
    \hline
    Parameter             & Value     & Error      &Parameter    && Value      &
Error   \\
    \hline \hline
    $m_{\tilde e_R}$      & 140.9 GeV & 0.05 GeV   & $\mu$       && 481.1 GeV  &
4.5 GeV \\
    $m_{\tilde e_L}$      & 190.4 GeV & 0.4 GeV    & $\tan\beta$ &&
   \phantom{0}10.0    & 1.0
  \\
    $m_{\tilde \nu_e}$    & 173.4 GeV & 1.3 GeV    & $M_{1/2}$   && 250.0 GeV  &
0.4 GeV \\
    $m_{\tilde \tau_1}$   & 104.2 GeV & 0.3 GeV    & $A_0$       & $-$&640.0 GeV &
13 GeV  \\
    $m_{\tilde \tau_2}$   & 187.8 GeV & 1.1 GeV    &             &&            &
        \\
    $m_{\tilde \nu_\tau}$ & 154.7 GeV & 1.6 GeV    &             &&            &
        \\
    \hline
  \end{tabular}
  \caption{\it Expected experimental errors for the determination of
  sfermion masses and other underlying parameters in the sample one-step SO(10)
  scenario. [The errors quoted correspond to 1$\sigma$.]}
           \label{tab:experror}
\end{table}
The masses of the charged sleptons can be measured with high
precision in slepton pair production at ILC \cite{a}, while the
sneutrino masses can be determined accurately from the decays of
charginos \cite{FPZ}. The Higgs masses in Eq.~(\ref{eq:higgsrel})
are expressed in terms of the parameters $\mu$ and $\tan\beta$,
which can be derived accurately from the analysis of chargino and
neutralino observables \cite{c1,c}. The pseudoscalar Higgs mass
$M_A$, on the  other hand, can be measured in associated Higgs
pair production \cite{HA}.
The direct determination of the $A$ parameters is difficult, {\it cf.}
Refs.~\cite{Bartl:1997yi,CMZ,Heinemeyer:2003ud}. However,
taking into account stop/sbottom mass measurements but leaving out
the sleptons of the third generation and the Higgs bosons
to bypass the (yet to be determined) parameters of the
right-handed neutrinos,
a global analysis including loop effects{\footnote{Based
on this reduced set of observables,
errors on $A_0$ of about 6 GeV have been predicted for SPS1a and SPS1a$'$
with $A_0$ values of $-100$ and $-300$ GeV, respectively. Fixing the
error on $A_0$ to 2\%, {\it i.e.} 12~GeV, for the present benchmark point
with large $A_0 = -640$ GeV
can therefore be considered as a conservative estimate.
Thanks go to P.~Bechtle and D.~Zerwas
for running Fittino and Sfitter, respectively,
for estimating the increase of the error
on $A_0$ in SPS1a and SPS1a$'$ for the
reduced set of experimental observables.}} leads to an accurate
determination of $A_0$ at the level of 2\%
\cite{spa,Bechtle:2004pc}.  The relevant
parameters are summarized in Tab.~\ref{tab:experror},
including  the estimated experimental errors.

\begin{figure}[t]
\centering
\includegraphics[clip,width=0.495\textwidth]{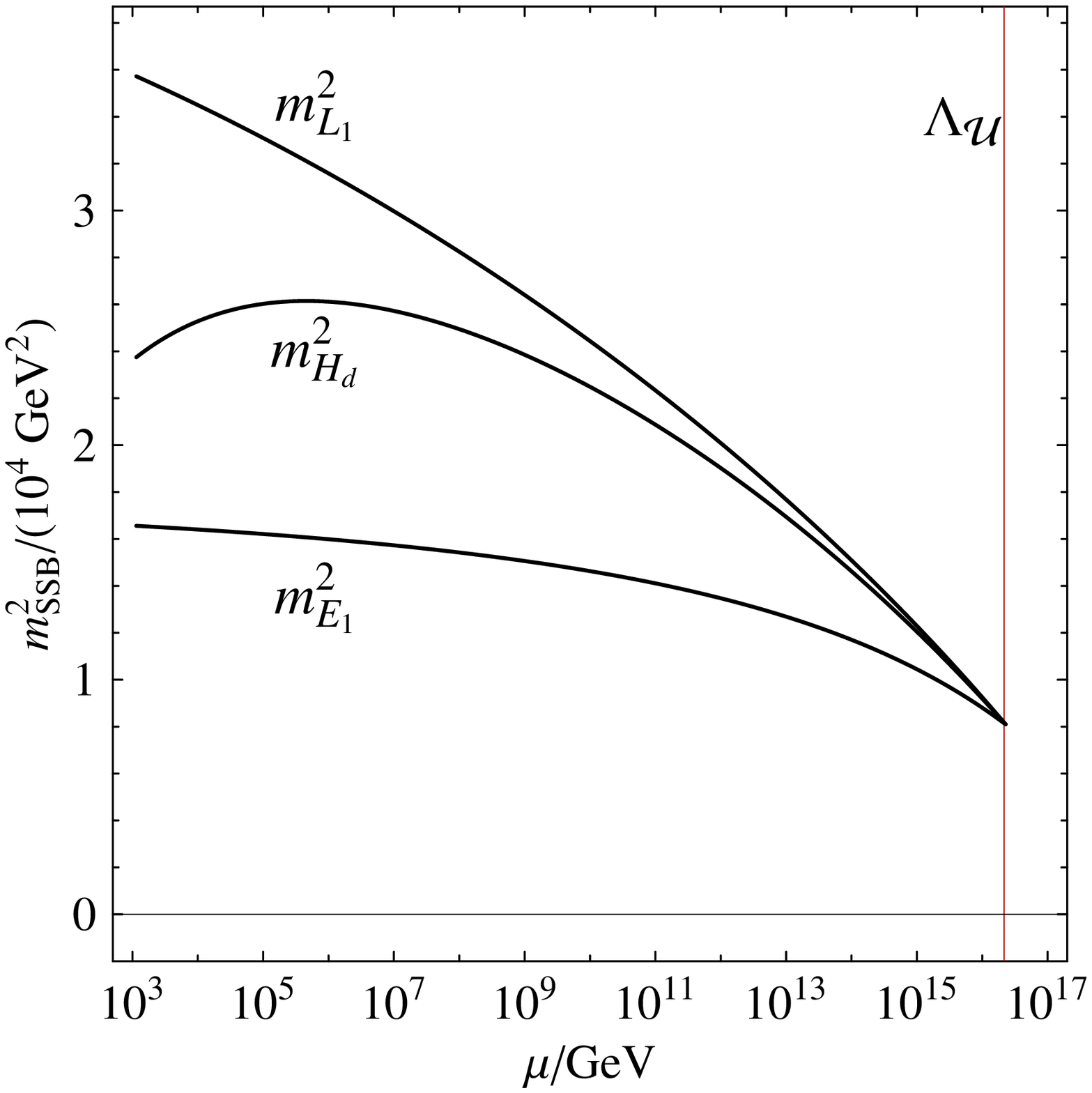}
\includegraphics[clip,width=0.495\textwidth]{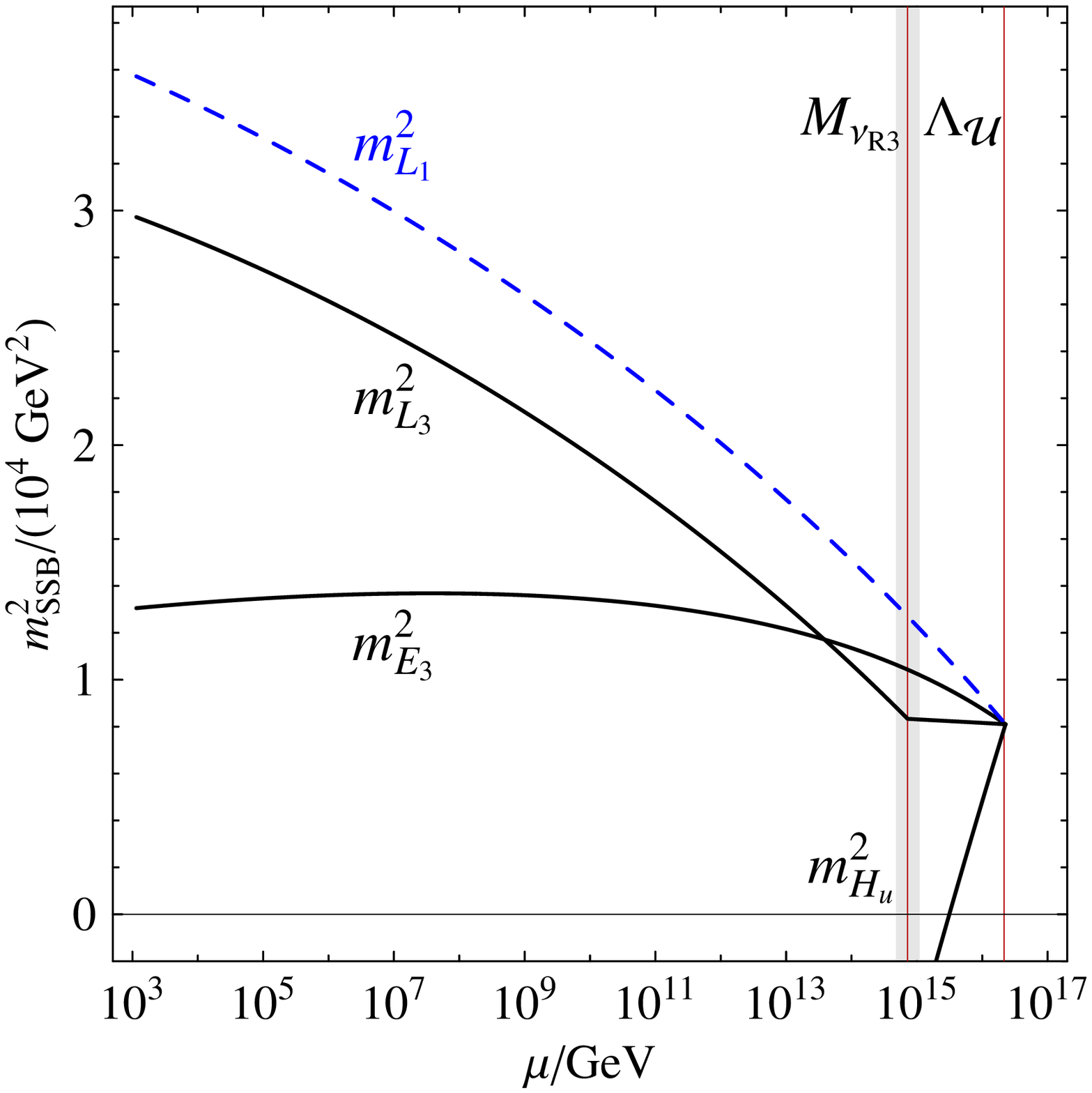}
     \caption{\it {\rm{Left:}} Evolution of the L,R slepton mass parameters of the first generation
              for one-step breaking with the D-term, merely for illustration, set to zero.
              {\rm{Right:}} Evolution of the first and third generation
                  L,R slepton and Higgs mass parameters [\(D_{\mathcal{U}}=0\)] for one-step breaking;
                  the kink in \(m^2_{L_3}\) is generated by the right-handed neutrino with mass
                  $M_{\nu_{R3}}$ close to $10^{15} \, {\rm GeV}$.}
     \label{fig:ScalarMassEvolutionGen1}
\end{figure}
The measurement of the slepton and sneutrino masses of the first
two generations allows us to extract the common sfermion parameter
$m_{16} = M_0$ as well as the D-term $D_{\mathcal{U}}$. The
relations are given in Eq.~\eqref{eq:RGfirstgen} in leading
logarithmic approximation for the RG running. Including the
complete one-loop and the leading two-loop corrections, the
evolution of the scalar mass parameters, $(m^2_{L_1}, m^2_{E1})$,
is displayed in Fig.~\ref{fig:ScalarMassEvolutionGen1}(left) for
the first two generations. The curves describe the evolution of
the universal L and R mass parameters with, merely for
illustration, the D-term set to zero.

\begin{table}[t]
  \centering
  \begin{tabular}{|l|c|c|}
    \hline
    Parameters in SO(10) $\to$ SM                & Ideal                      & Error     \\
    \hline\hline
    unification scale  \(\Lambda_{\mathcal U}\)         &  $2.16 \cdot 10^{16}$~GeV  & $0.02 \cdot 10^{16}$~GeV \\
    matter scalar mass $M_0$                 &  90~GeV                    & 0.25~GeV  \\
    GUT D-term         $\sqrt{-D_{\mathcal{U}}}$         &  30~GeV                    & 0.9~GeV   \\
    \hline
    heaviest R-neutrino mass $M_{\nu_{R 3}}$ & $7.2 \cdot 10^{14}$~GeV    &
        $[4.8,11]\cdot 10^{14}$~GeV \\
    lightest neutrino mass \(m_{\nu_1}\)      & $3.5 \cdot 10^{-3}$~eV    &
        $[1.6,6.7]\cdot 10^{-3}$~eV \\
    \hline
  \end{tabular}
  \caption{\it Reconstruction of high-scale SO(10) parameters
               in one-step SO(10) $\to$ SM breaking, and masses
               of the heavy 3-generation R-neutrino and the lightest neutrino
               [The errors quoted correspond to 1$\sigma$].}
           \label{tab:ParameterDetermination}
\end{table}
With the estimated errors in Tab.~\ref{tab:experror}, the
high-scale parameters can be calculated, by combining the slepton
and Higgs sectors, as shown in
Tab.~\ref{tab:ParameterDetermination}. The RG evolution equations
are evaluated to 2-loop order by means of the SPheno program \cite{Spheno}.
The table indicates that the high-scale parameters $M_0$ and $D_{\mathcal{U}}$,
driven by the slepton analysis,
can be reconstructed at per-mill to per-cent accuracy.

\begin{figure}[t]
\centering
\includegraphics[clip,width=0.77\textwidth]{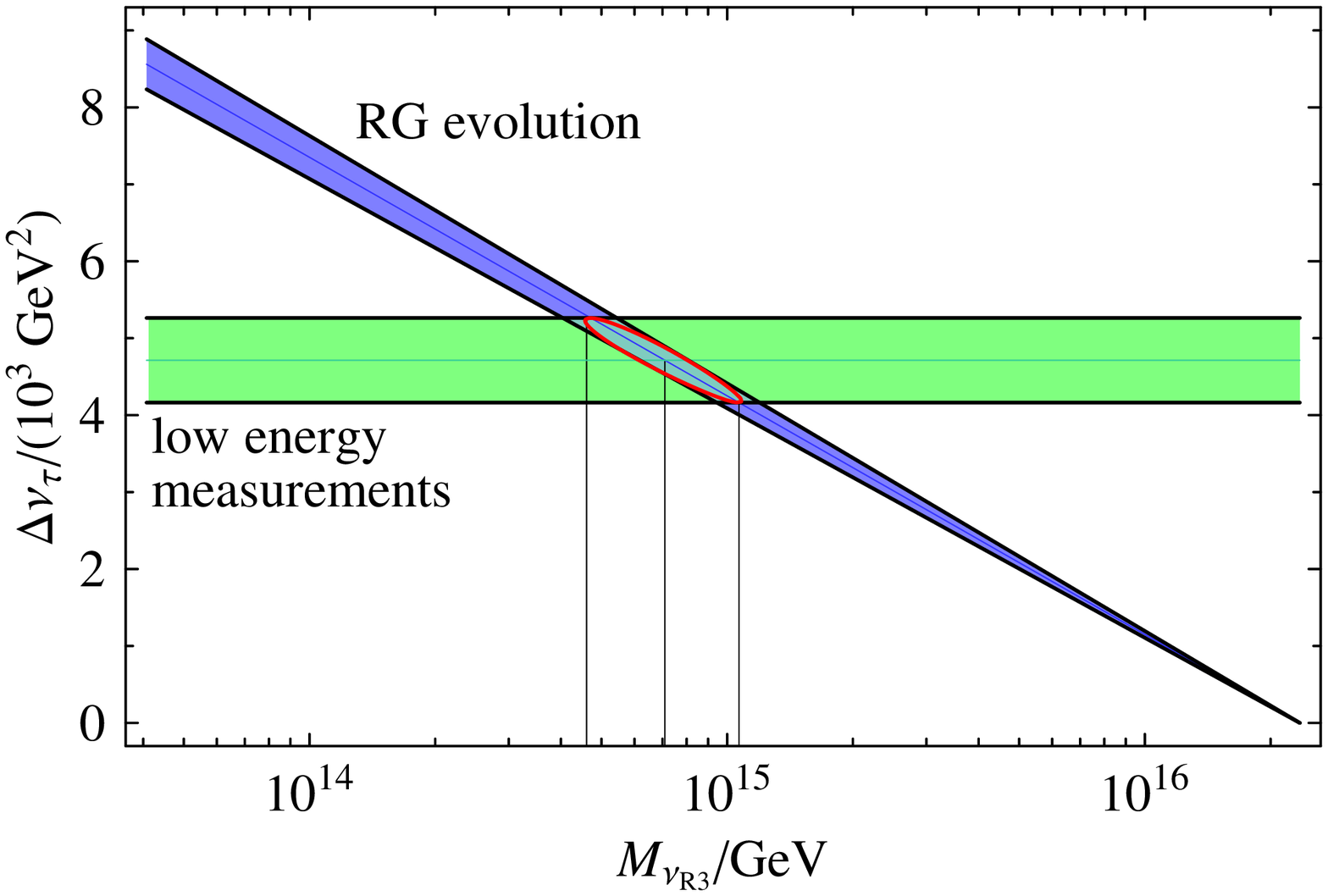}
\vspace{-1.5em}
     \caption{\it Shift \(\Delta_{\nu_\tau}\) of the third generation L slepton mass
     parameter generated by loops involving heavy neutrino {\rm{R}}-superfields. The
     blue wedge
     corresponds to the prediction Eq.~(\ref{eq:DeltaNuTau}) of the renormalization group [RG],
     whereas the green band is determined by low-energy mass measurements.}
     \label{fig:Wedge}
\end{figure}
The right-handed neutrino affects the evolution of the L mass
parameter $m^2_{L_3}$ in the third generation and the Higgs
parameter $m^2_{H_u}$. [Note that $m^2_{H_u} + \mu^2$ turns
negative only at the small scale $Q \simeq 350$~GeV.] The
characteristic difference in the evolution between $m^2_{L_3}$ and
$m^2_{L_1}$ is exemplified in
Fig.~\ref{fig:ScalarMassEvolutionGen1}(right) for a right-handed
neutrino mass $M_{\nu_{R3}}$ of $7.2 \cdot 10^{14}$ GeV. ~From the
universality point $\Lambda_{\mathcal U}$, where $m_{L_3}$ and
$m_{L_1}$ are equal, down to the kink at $M_{\nu_{R3}}$ the
evolution of $m_{L_3}$ is affected by the right-handed neutrino,
$m_{L_1}$ however is not. Below the kink the difference between
$m_{L_3}$ and $m_{L_1}$ is reduced to the standard loop correction
$\Delta_\tau$ and the small $\tau$ mass. The position of the kink
at $M_{\nu_{R3}}$ can be derived from the measured slepton and
sneutrino masses if the scalar mass parameters are universal at
the unification scale. Neglecting the small term $m^2_\tau$, the
right-handed neutrino mass is fixed by the intersection of the
parameter $\Delta_{\nu_\tau}$ as a function of $M_{\nu_{R3}}$ with
the measured value extracted from the slepton masses:
\begin{eqnarray}
\Delta_{\nu_\tau} &=& \frac{1}{2} [m^2_{{\tilde{e}}_L} +
m^2_{{\tilde{e}}_R} - 3 m^2_{{\tilde\nu}_{e L}}] - \frac{1}{2}
[m^2_{{\tilde\tau}_1} + m^2_{{\tilde\tau}_2} - 3
m^2_{{\tilde\nu}_{\tau L}}]    \nonumber \\
                  &=& \frac{m_t^2(\Lambda_{\mathcal U})}{8 \pi^2
                      v_u^2}\left(3M_0^2+A_0^2\right)\log\frac{\Lambda_{\mathcal U}^2}{M_{\nu_{R3}}^2}  \,.
\end{eqnarray}

In practice the separate evaluation of the charged slepton and sneutrino
mass differences,
\begin{eqnarray}
\Delta_{\nu_\tau} &=&
        m^2_{\tilde\nu_{e L}} - m^2_{\tilde\nu_{\tau L}} - \Delta_\tau \nonumber \\
                  &=& m^2_{\tilde e_L} + m^2_{\tilde e_R}
      - m^2_{\tilde \tau_L} - m^2_{\tilde \tau_R} - 3\Delta_\tau
\end{eqnarray}
proves useful to obtain a more precise value,
$\Delta_{\nu_\tau}^{exp} = (4.7 \pm 0.5) \cdot 10^{3}$ GeV$^2$.
As a result, for the parameters of the LR extended point
SPS1a$'$ introduced earlier, the right-handed neutrino mass of the
third generation is estimated in the margin
\begin{equation}
  M_{\nu R_3} = 10^{14.86 \pm 0.19} {\rm{\ GeV}} =
  7.2^{+4.0}_{-2.5} \cdot 10^{14} {\rm{\ GeV}}
\end{equation}
as evident from Fig.~\ref{fig:Wedge}.
Thus, the effect of the heavy $\nu_{R3}$ mass can indeed be traced back
from measured slepton
masses in universal supersymmetric theories.

Based on this estimate of $M_{\nu R_3}$, the seesaw mechanism
determines the value
of the lightest neutrino mass ({\it cf.} Fig.~\ref{fig:Mivsm1}) to
\begin{equation}
  m_{\nu_1} = 10^{-2.51 \pm 0.32} {\rm{\ eV}}
  = 3.1^{+3.5}_{-1.5} \cdot 10^{-3} {\rm{\ eV}}.
\end{equation}
The second lightest neutrino mass $m_{\nu_{2}} \simeq (1.0\pm 0.1)
\cdot 10^{-2}$ eV is about three times larger while the third
neutrino mass $m_{\nu_{3}} \simeq 5 \cdot 10^{-2}$ eV coincides
with the mass difference $\Delta m_{13}$ measured in the neutrino
oscillation experiments.

Thus, the combination of SO(10) symmetry and seesaw mechanism
leads, besides the high-scale universal SUSY parameters, to the
determination of the heavy Majorana mass \(M_{\nu_{R 3}}\) of the
third generation and, in a consecutive step, to the value of the
lightest neutrino mass \(m_{\nu_1}\) in hierarchical theories.
This conclusion has been derived for scenarios in which threshold effects
at $\Lambda_{\mathcal U}$ and the mechanisms for solving the
doublet-triplet splitting problem in the Higgs sector {\it etc.}
do not have a significant impact on the evolution of the scalar
mass parameters. The analysis however may serve
as a generic paradigm which could be adjusted correspondingly if
the structure of the high-scale scenario demanded a more complex
extension.

\section{Two-Step SO(10) \(\to\) SU(5) \(\to\) SM Breaking}\label{sec:twostep}

One possible move in the direction of higher complexity is
the extension of the analysis to a simple model of 2-step
breaking. If the grand unification symmetry SO(10) is broken in
two steps down to the Standard Model, {\it cf.}
Fig.~\ref{fig:BreakingDiagrams}(right), the evolution between the
Terascale and the SU(5) scale \(\Lambda_{\mathcal U}\) is
determined by measured parameters: gauge couplings, scalar mass
parameters and neutrino parameters. However, the evolution between
the SU(5) scale \(\Lambda_{\mathcal U}\) and the SO(10) scale
\(\Lambda_{\mathcal{O}}\) depends on the high-scale physics
scenario, comprising new matter, gauge and Higgs fields, and
potentially effective elements of gravity interactions.

In the present approach
we restrict these degrees of freedom to the $\{24\}$ Higgs field
which breaks the SU(5) gauge symmetry.
In the same spirit as in the previous section,
other degrees of freedom and/or
additional mechanisms generating the doublet-triplet splitting
and prolonging the proton lifetime, for example, are assumed to be weakly
coupled to the scalar mass parameters below $\Lambda_{\mathcal{O}}$.
The numerical results in this two-step analysis are
less robust than in the previous one-step analysis
since they depend explicitly on the physics scenario
at the high scales beyond $> \Lambda_U$.
Nevertheless this study can serve as an illustration
for the potential of parametric analyses
if the fundamental structure of a more
complex scenario is theoretically pre-determined. \\[-2mm]

{\it a) Evolution Terascale $\to \Lambda_{\mathcal U}$} \\[3mm]
Based on the standard SU(5) decomposition, the scalar masses
evolve from the electroweak scale to the SU(5) scale
$\Lambda_{\mathcal U}$ according to the rules
\begin{eqnarray}
  m_{\tilde e_R}^2&=&
  m_{10,1}^2   + \alpha_R M_{\mathcal{U},1/2}^2 - \tfrac{6}{5} S' -2s_W^2D_{EW}  \nonumber \\
  m_{\tilde e_L}^2    &=&
  m_{\bar{5},1}^2 + \alpha_L M_{\mathcal{U},1/2}^2 +  \tfrac{3}{5} S' -c_{2W} D_{EW}  \nonumber \\
  m_{\tilde\nu_{eL}}^2 &=&
  m_{\bar{5},1}^2 + \alpha_L M_{\mathcal{U},1/2}^2 +  \tfrac{3}{5} S'+D_{EW}
  \label{eq:ScalarGenOneTwo}
\end{eqnarray}
for the matter fields of the first two generations while the masses of the third generation,
\begin{eqnarray}
  m_{\tilde \tau_R}^2 &=&
  m_{10,3}^2 + \alpha_R M_{\mathcal{U},1/2}^2 - \tfrac{6}{5} S'   -2s_W^2D_{EW} +m_\tau^2  -2\Delta_\tau   \nonumber \\
  m_{\tilde \tau_L}^2  &=&
  m_{\bar{5},3}^2 + \alpha_L M_{\mathcal{U},1/2}^2 +  \tfrac{3}{5} S' - c_{2W} D_{EW} + m_\tau^2
                                                            - \Delta_\tau - \Delta_{\nu_\tau}    \nonumber \\
  m_{\tilde\nu_{\tau L}}^2 &=&
  m_{\bar{5},3}^2 + \alpha_L M_{\mathcal{U},1/2}^2 +  \tfrac{3}{5} S'+D_{EW} -  \Delta_\tau - \Delta_{\nu_\tau}
  \,,
\end{eqnarray}
are affected in addition by the Yukawa interactions in the
same way as Eqs.~(\ref{eq:DeltaTau},\ref{eq:DeltaNuTau}). The indices $1,3$
next to the SU(5) multiplet characteristics denote the generation numbers. The
evolution of the Higgs mass parameters read correspondingly
\begin{eqnarray}
  m^2_{H_d} &=& m_{\bar{5}_1}^2  + \alpha_L M_{\mathcal{U},1/2}^2 +  \tfrac{3}{5} S' - \Delta_\tau
     - 3 \Delta_b   \label{eq:HiggsDown}
\\
  m^2_{H_u} &=& m_{5_2}^2  + \alpha_L M_{\mathcal{U},1/2}^2  -  \tfrac{3}{5} S' - \Delta_{\nu_\tau}
     - 3 \Delta_t   \,.
\end{eqnarray}
Since SU(5) and SU(3)$\times$SU(2)$\times$U(1) are both rank-4
symmetry groups, no GUT D-term is induced in this symmetry
breaking
step.

The $\Delta$'s are given by the loop corrections,
\begin{eqnarray}
  \Delta_\tau        &=& \frac{m^2_\tau(\Lambda_{\mathcal U})}{8 \pi^2 v^2_d}
     \left(m_{10,3}^2+m_{\bar{5},3}^2+m_{\bar{5}_1}^2+A_{5,3}^2\right)\log\frac{\Lambda^2_{\mathcal U}}{\tilde{M}^2}\\
  \Delta_{\nu_\tau}  &=& \frac{m^2_t(\Lambda_{\mathcal U})}{8\pi^2 v_u^2}\left(m_{1,3}^2+
m_{\bar{5},3}^2+m_{5_2}^2+A_{1,3}^2\right)\log\frac{\Lambda^2_{\mathcal U}}{M^2_{\nu_{R3}}}
\end{eqnarray}
and
\begin{eqnarray}
  \Delta_b &=& \frac{m^2_b(\Lambda_{\mathcal U})}{8\pi^2 v^2_d}\left(m_{10,3}^2+m_{\bar{5},3}^2+
    m_{\bar{5}_1}^2+A_{5,3}^2\right)\log\frac{\Lambda^2_{\mathcal U}}{\tilde{M}^2}\\
  \Delta_t &=& \frac{m_t^2(\Lambda_{\mathcal U})}{8\pi^2 v^2_u}\left(2m_{10,3}^2+m_{5_2}^2
                      +A_{10,3}^2\right)\log\frac{\Lambda^2_{\mathcal U}}{\tilde{M}^2} \,,
\end{eqnarray}
while the contributions to the $S'$ parameter trace back to the
non-universal Higgs mass parameters at the scale
$\Lambda_{\mathcal U}$,
\begin{equation}
  S' =
  \frac{\alpha_1(\tilde{M})}{\alpha_1(\Lambda_{\mathcal U})}\left(m_{5_2}^2-m_{\bar{5}_1}^2\right).
\end{equation}
$M_{\mathcal{U},1/2}$ is the universal gaugino mass parameter at the SU(5)
scale $\Lambda_{\mathcal U}$.

\subsubsection*{\it b) Evolution $\Lambda_{\mathcal U} \to \Lambda_{\mathcal{O}}$:}
The subsequent evolution from the SU(5) breaking scale
$\Lambda_{\mathcal U}$ to the SO(10) breaking scale
$\Lambda_{\mathcal{O}}$ unifies the mass parameters $m_{10}$,
$m_{\bar{5}}$ and $m_1$ to $m_{16}$,
\begin{eqnarray}
  m_{10,1}^2       &=& M^2_0  + D_{\mathcal{O}} + \alpha'_R M_{1/2}^2    \nonumber\\
  m_{\bar{5},1}^2  &=& M^2_0 -3 D_{\mathcal{O}} + \alpha'_L M_{1/2}^2
\end{eqnarray}
and
\begin{eqnarray}
  m_{10,3}^2       &=& M^2_0  + D_{\mathcal{O}} + \alpha'_R M_{1/2}^2 -3\Delta'_t - 2\Delta'_b  \nonumber\\
  m_{\bar{5},3}^2  &=& M^2_0 -3 D_{\mathcal{O}} + \alpha'_L M_{1/2}^2 - \Delta'_{\nu_\tau} - 4\Delta'_b   \nonumber\\
  m_{1,3}^2        &=& M^2_0 +5 D_{\mathcal{O}} -5\Delta'_{\nu_\tau}                                    \,.
\end{eqnarray}
The Higgs parameters transform according to
\begin{eqnarray}
  m_{\bar{5}_1}^2  &=& M^2_0 + 2 D_{\mathcal{O}} + \alpha'_L M_{1/2}^2
        - 4 \Delta_b'  - \tfrac{24}{5} \Delta'_\lambda      \,,             \nonumber \\
  m_{5_2}^2  &=& M^2_0 - 2 D_{\mathcal{O}} + \alpha'_L M_{1/2}^2
        -  \Delta_{\nu_\tau}'  - 3 \Delta_t' - \tfrac{24}{5} \Delta'_\lambda \,.
\end{eqnarray}

The D-term associated with the breaking of the rank-5 SO(10) to
the rank-4 SU(5) symmetry is given again by a relation analogous
to Eq.~(\ref{eq:Dterm}) for a Higgs-16 field, for example,
responsible for the symmetry breaking of SO(10) $\to$ SU(5). The
universal gaugino mass parameter $M_{1/2}$ at
$\Lambda_{\mathcal{O}}$ is related to the universal parameter
$M_{\mathcal{U},1/2}$ at $\Lambda_{\mathcal U}$ by
\begin{equation}
  M_{1/2} = \frac{\alpha_{\mathcal{O}}}{\alpha_{\mathcal U}} \, M_{\mathcal{U},1/2} \,.
\end{equation}
The coefficients
$\alpha'_{L,R}$
are given by
\begin{eqnarray}
\alpha'_{L} &=& \frac{3}{2} \alpha'_{R}  =\frac{36}{5 b_5} \left( 1
 - \left[{1+\frac{g^2_{SO(10)}}{16 \pi^2} b_5 \log
               \frac{\Lambda^2_{\mathcal{O}}}{\Lambda_{\mathcal U}^2}}\right]^{-2}
               \right),
\end{eqnarray}
with $b_5 = -3$ in the minimal SU(5) model including the Higgs
$\{24\}$ representation, {\it cf.}
Fig.~\ref{fig:BreakingDiagrams}(right):
\begin{eqnarray}
   b_5 &=& b_5(\mbox{matter}) + b_5(\mbox{gauge}) + b_5(\mbox{Higgs})   \nonumber \\
   b_5(\mbox{matter}) &=& 6 \,;\;\; b_5(\mbox{gauge}) = -15\,;\nonumber \\
 b_5(\mbox{Higgs}\;\{24\}) &=& 5
  \,;\;\; b_5(\mbox{Higgs}\;\{5\}+\{\bar{5}\}) = 1.
\end{eqnarray}
The $\Delta'$ coefficients,
\begin{eqnarray}
\Delta_t' &\approx&  \Delta_{\nu_\tau}' \approx
 \frac{m_t^2(\Lambda_{\mathcal{O}})}{8\pi^2
        v_u^2}\left(3m_{16}^2+A_0^2\right)\log\frac{\Lambda^2_{\mathcal{O}}}{\Lambda_{\mathcal U}^2}
 \label{eq:DeltaTp}\\
\Delta_b' &\approx&  \frac{m_b^2(\Lambda_{\mathcal{O}})}{8\pi^2
       v_d^2}\left(3m_{16}^2+A_0^2\right)\log\frac{\Lambda^2_{\mathcal{O}}}{\Lambda_{\mathcal U}^2} \, ,
 \label{eq:DeltaBp}
\end{eqnarray}
are (s)quark and (s)lepton loop contributions to the transport
from $\Lambda_{\mathcal U}$ to $\Lambda_{\mathcal{O}}$.
The shift $\Delta'_\lambda$, accounting for contributions
involving the heavy $\{24\}$ Higgs field which couples to the $\{5\}$
and $\{\bar{5}\}$ Higgs fields, is small in general; moreover,
it can be neglected since the Higgs sector, in effect, plays
a minor role in the analysis.

The evolution of the gauge coupling from $\Lambda_{\mathcal U}$ to
$\Lambda_{\mathcal{O}}$,
\begin{equation}
\alpha(\Lambda_{\mathcal{O}}) = \frac{\alpha(\Lambda_{\mathcal{U}})}{1- \frac{\alpha(\Lambda_{\mathcal{U}})}
               {2 \pi} b_5 \log{\frac{\Lambda_{\mathcal{O}}^2}{\Lambda_{\mathcal U}^2}}} \, ,
\end{equation}
is affected by the Higgs-24 field which breaks SU(5) $\to$ SM.

Finally, the shift of the $A$ parameters between
$\Lambda_{\mathcal U}$ and $\Lambda_{\mathcal{O}}$ is small:
\begin{eqnarray}
A_{5,3} &\simeq& 0.97 A_0 - 0.15 M_{1/2} \\
A_{10,3}  &\simeq& 0.91 A_0 - 0.16 M_{1/2} \\
A_{1,3}  &\simeq&  0.89 A_0 - 0.10 M_{1/2}
\end{eqnarray}
for the reference point defined below.

This 2-step breaking system has been analyzed quantitatively for a
parameter set closely related to the previous
example:
\begin{equation}
\begin{array}{rclrcl}
   M_0 &=& 100 {\rm{\ GeV}} \hspace{1.5cm}
   & \tan\beta &=& 20  \\
   M_{1/2}       &=& 230 {\rm{\ GeV}} & {\rm sgn}(\mu)     &=& +         \\
   A_0           &=& -425 {\rm{\ GeV}}
\end{array}
\end{equation}
and
\vspace*{-3mm}
\begin{eqnarray}
   \Lambda_{\mathcal{O}} &=& 5 \cdot 10^{17} {\rm{\ GeV}}             \nonumber \\
   D_{\mathcal{O}}       &=&  0.9 \cdot 10^3 {\rm{\ GeV}}^2          \nonumber \\
   M_{\nu_{R_3}}         &=& 6.3 \cdot 10^{14} {\rm{\ GeV}}          \, .
\end{eqnarray}

The scenario{\footnote{Note that this high-scale point
is outside the parameter range
investigated in Ref.~\cite{YM}; we thank Y. Mambrini
for a clarifying comparison.}}
is compatible with values of the low-energy
and cosmological data specified earlier: $\alpha_s(M_Z)=0.119$,
${\sin^2\theta}^{lept}_{eff}= 0.23140$,
BR$(b \to s \gamma) = 3.11 \cdot 10^{-4}$,
$\Delta a_\mu = 25.5 \cdot 10^{-10}$, and $\Omega h_0^2 = 0.094 $.

\begin{table}[!t]
  \centering
  \begin{tabular}{|l|l|l||l|l@{}l|l|}
    \hline
    Parameter             & Value     & Error      &Parameter    && Value      &
Error   \\
    \hline \hline
    $m_{\tilde e_R}$      & 192.2 GeV & 0.07 GeV   & $\mu$       && 429.5 GeV  &
5.0 GeV \\
    $m_{\tilde e_L}$      & 217.8 GeV & 0.5 GeV    & $\tan\beta$ && \phantom{0}20.0    & 5.0
 \\
    $m_{\tilde \nu_e}$    & 202.7 GeV & 0.7 GeV    & $M_{1/2}$   && 230.0 GeV  &
0.4 GeV \\
    $m_{\tilde \tau_1}$   & 101.9 GeV & 0.3 GeV    & $A_0$       & $-$&425.0 GeV &
8.5 GeV  \\
    $m_{\tilde \tau_2}$   & 219.4 GeV & 1.2 GeV    &             &&            &
        \\
    $m_{\tilde \nu_\tau}$ & 178.0 GeV & 1.0 GeV    &             &&            &
        \\
    \hline
  \end{tabular}
  \caption{\it Expected experimental errors for the determination of sfermion
  masses and other underlying parameters in the two-step SO(10) $\to$ SU(5)
  $\to$ SM scenario. [The errors quoted correspond to 1$\sigma$.]}
           \label{tab:experror2}
\end{table}
The analysis is performed in the same way as before, with the
measurement errors extrapolated from existing studies to this
scenario as listed in Tab.~\ref{tab:experror2}. The error on $A_0$
can be estimated again to be 2\% according to the guideline
followed in the one-step analysis, as the evolution from
$\Lambda_{\mathcal U}$ to $\Lambda_{\mathcal{O}}$ has little
impact on the $A$ parameters. Though uncertainties due to the
shift of the universality scale are expected to be small, the
analysis has been repeated for an error increased to 5\%,
nevertheless. The final results are affected only slightly.

\begin{figure}[!t]
\centering
\includegraphics[clip,width=0.77\textwidth]{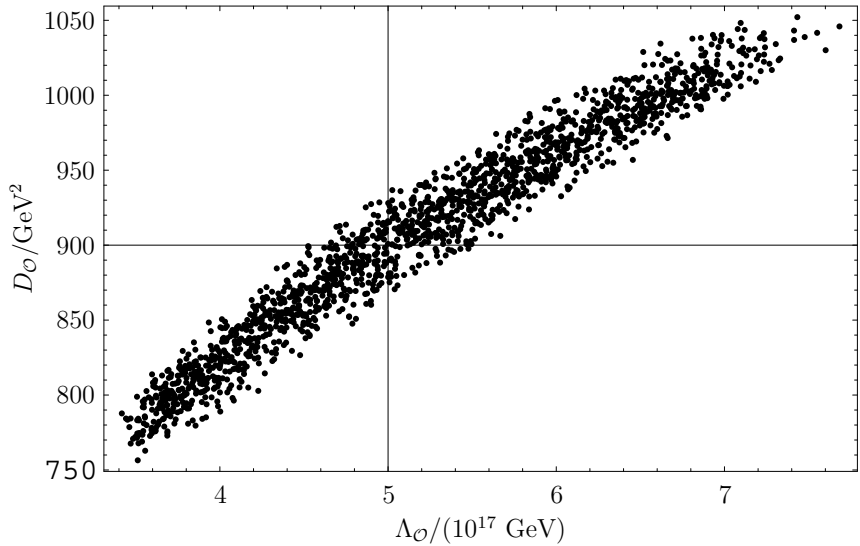}
     \caption{\it Correlation between the SO(10)
unification scale $\Lambda_{\mathcal{O}}$ and the D-term
$D_{\mathcal{O}}$ in the analysis of the two-step SO(10) $\to$
SU(5)
  $\to$ SM scenario. The scatter plot represents a $\chi^2$ analysis based
  on the extrapolation of the expected experimental errors from the weak scale
  to the unification scale.}
     \label{fig:corrLODO}
\end{figure}
As naturally expected, a strong correlation between the D-term
$D_{\mathcal{O}}$ and the SO(10) unification scale
$\Lambda_{\mathcal{O}}$ is observed, {\it cf.}
Fig.~\ref{fig:corrLODO}. Since $D_{\mathcal{O}}$ is predicted
theoretically when the SO(10)~$\to$~SU(5) breaking mechanism is
devised, $D_{\mathcal{O}}$ is fixed in the following analysis
while $\Lambda_{\mathcal{O}}$ is allowed to float. Results of the
reverse analysis are briefly summarized for completeness.

\begin{figure}[!t]
\centering
\includegraphics[clip,width=0.495\textwidth]{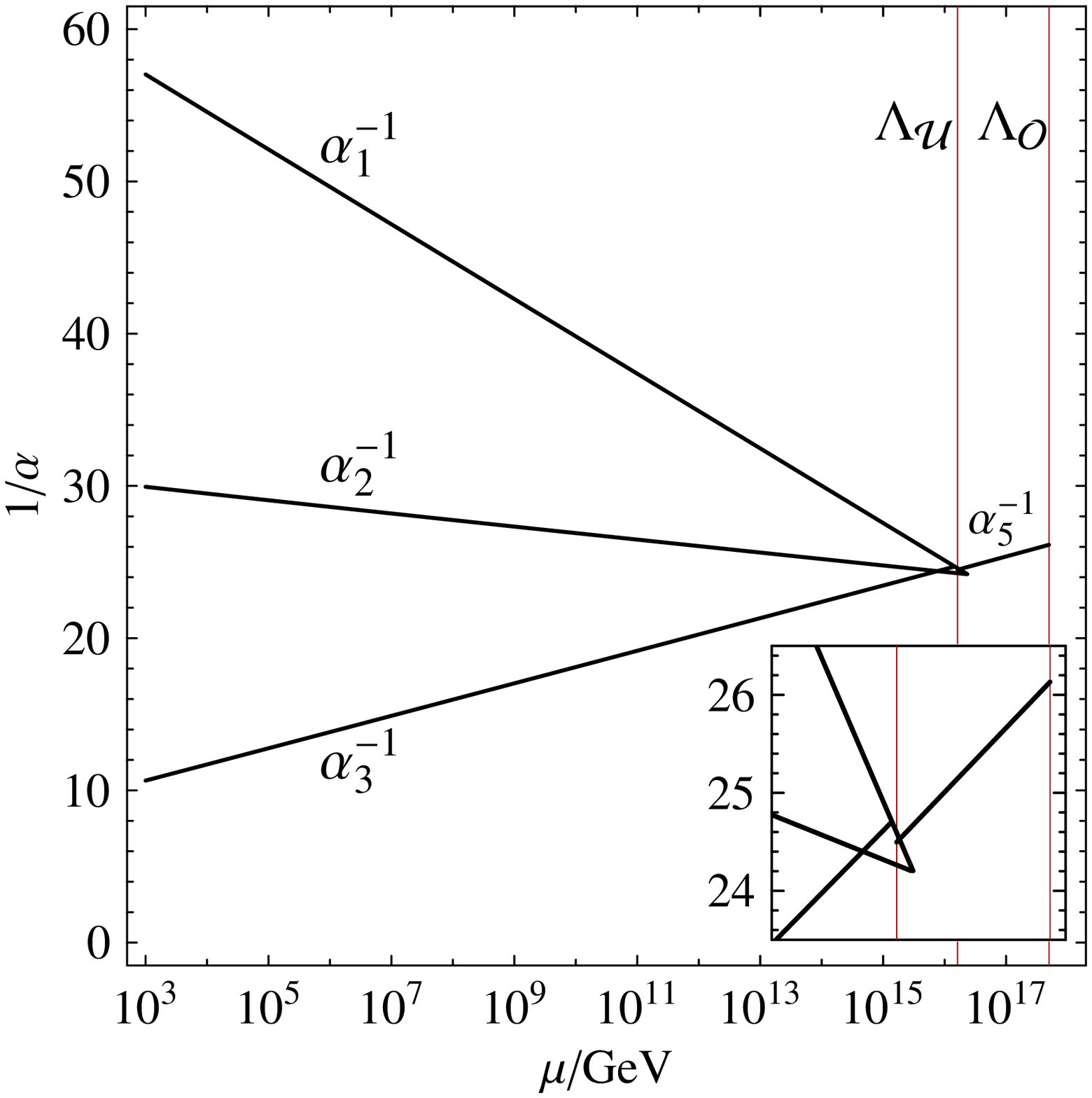}
\includegraphics[clip,width=0.495\textwidth]{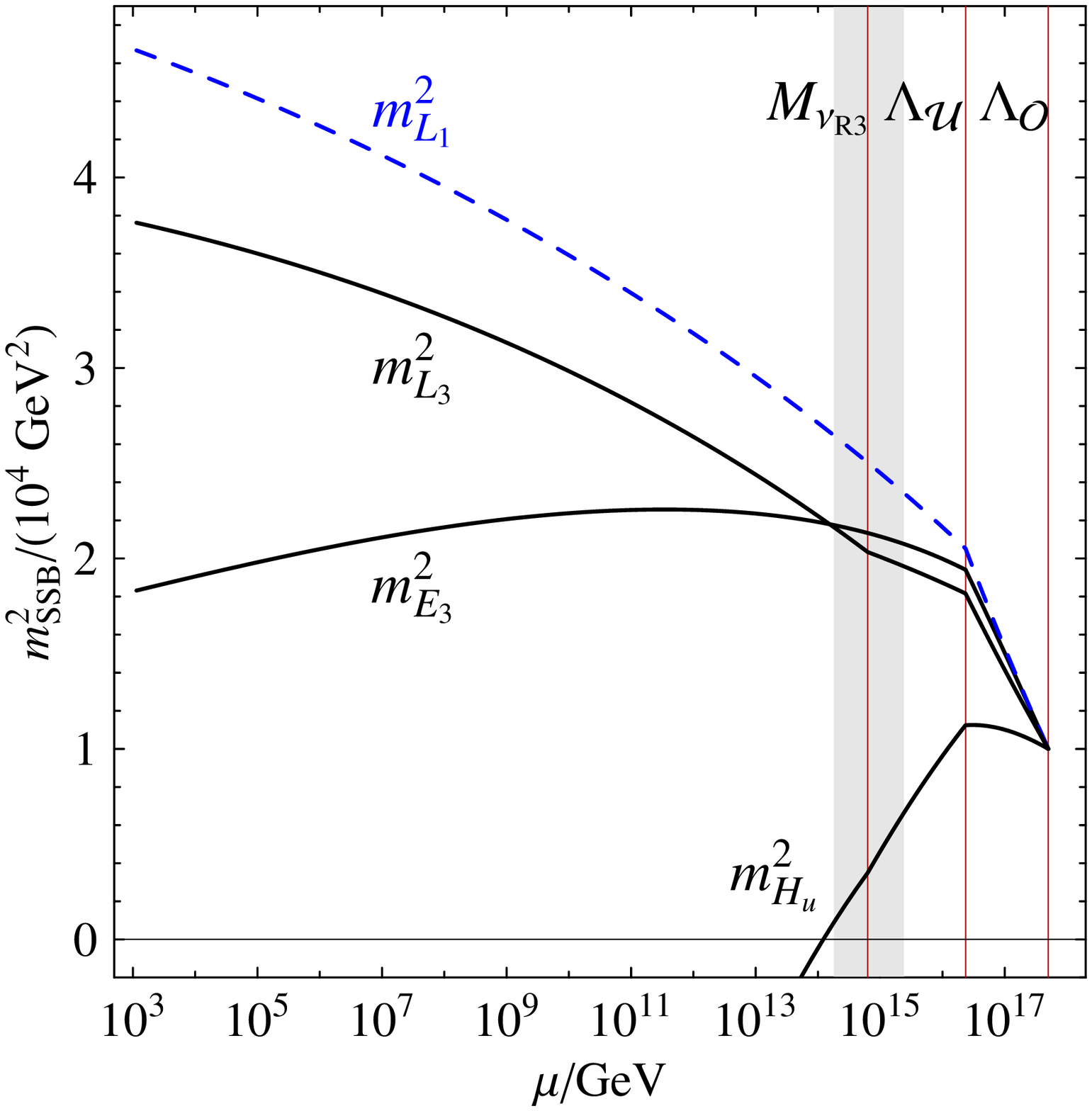}
     \caption{\it {\rm Left:} Evolution of the gauge couplings in 2-step
                              $\,${\rm SO(10)} $\to$ {\rm SU(5)} $\to$ {\rm SM}
                              symmetry breaking for the {\rm SU(5)} Higgs
                              representation {\rm \{24\}}; {\rm Right:}
                              Evolution of scalar matter and down-type Higgs
                              mass parameters, $D_{\mathcal{O}} = 0$ chosen
                              for illustration.
             }
     \label{fig:ScalarMassEvolutionSU5}
\end{figure}
Inspecting the solutions of the evolution equations, and
Fig.~\ref{fig:ScalarMassEvolutionSU5}, a simple picture emerges
for the reconstruction of the high-scale theory:

{\bf (i)} The SU(5) scale $\Lambda_{\mathcal U}$ can be derived
from the unification point of the strong, electromagnetic and weak
couplings, {\it cf.} Fig.~\ref{fig:ScalarMassEvolutionSU5}(left),
continuing to $\Lambda_{\mathcal{O}}$ by including the $\{24\}$
Higgs representation for SU(5) breaking;

{\bf (ii)} The SU(5) mass parameters $m_{10},\,m_{\bar{5}}$ are
determined by the measured slepton masses of the first two
generations, and the down-type scalar Higgs mass parameter
analogously. Noticeable deviations from universality are observed
for the mass parameters at the intermediate SU(5) GUT scale
$\Lambda_{\mathcal U}$. [The evolution equations,
Eqs.~(\ref{eq:ScalarGenOneTwo}) and (\ref{eq:HiggsDown}), depend
on $S'$, which involves the difference between the down- and
up-type Higgs mass parameters, $m_{{\bar5}_1}$ $m_{5_2}$. The
corrections from the small difference of these parameters have
been properly taken into account after performing the next step in
the analysis based on the universality of the soft supersymmetry
breaking parameters at $\Lambda_{\mathcal{O}}$];

{\bf (iii)} The triple meeting point of the SU(5) parameters of
the first two generations and the down-type Higgs in the evolution
from the SU(5) to the SO(10) grand unification scale determines
the SO(10) mass parameter $M_0$ and the GUT scale
$\Lambda_{\mathcal{O}}$, {\it cf.}
Fig.~\ref{fig:ScalarMassEvolutionSU5}(right)%
;

{\bf (iv)} By matching the evolution of the slepton mass
parameters in the third generation and the up-type Higgs with
$M_0$, the value of the heavy R-neutrino mass $M_{\nu_{R 3}}$ in
the seesaw mechanism can be estimated. The mass $8.6 \cdot
10^{13}$ GeV $< M_{\nu_{R3}} < 5.0 \cdot 10^{15}$ GeV can only be
estimated very roughly in this specific two-step scenario. This is
not primarily due to the increased complexity of the 2-step
structure, but rather due to the smaller value of \(A_0\) compared
to the 1-step scenario; this leads to a reduced slope in the RG
evolution of $\Delta_{\nu_\tau}$ ({\it cf.} Fig.~\ref{fig:Wedge})
so that the uncertainty rises in \(M_{\nu_{R3}}\). [Heavier slepton
masses, however, would also reverse this tendency and reduce the
error.] The uncertainty is only slightly larger, $8.0 \cdot
10^{13}$ GeV $< M_{\nu_{R3}} < 6.0 \cdot 10^{15}$ GeV, if the
error on the $A_0$ parameter is increased to 5\% as indicated
before. If the integrated luminosities are raised to 1 ab$^{-1}$
both at (S)LHC and ILC, the estimate of the R-sneutrino mass narrows
down to $1.8 \cdot 10^{14}$ GeV $< M_{\nu_{R3}} < 2.4 \cdot
10^{15}$ GeV, {\it i.e.} almost by an order of magnitude. These
estimates point clearly to a value in the area just below the
SU(5) unification scale.

\begin{table}[!h]
  \centering
  \begin{tabular}{|l|c|c|c|}
    \hline
    Parameter    & Mass/Scale
        & Error for 500 fb$^{-1}$ &  Error for 1000 fb$^{-1}$\\
    \hline\hline
    SU(5) unification scale  \(\Lambda_{\mathcal U}\) & 2.35$\cdot 10^{16}$~GeV
        & 0.02$\cdot 10^{16}$~GeV & 0.02$\cdot 10^{16}$~GeV \\
    matter scalar mass \(m_{10,1}\)        & 163.3~GeV
        & 0.14~GeV  & 0.10~GeV \\
    matter scalar mass \(m_{\bar{5},1}\)   & 133.5~GeV
        & 0.6~GeV & 0.45~GeV \\
    matter scalar mass \(m_{10,3}\)        & 142.2~GeV
        & 1.2~GeV & 0.85~GeV \\
    matter scalar mass \(m_{\bar{5},3}\)   & 124.2~GeV
        & 0.45~GeV & 0.4~GeV \\
    Higgs scalar mass  \(m_{\bar{5}_1}\)   & 127.8~GeV
        & 0.5~GeV & 0.35~GeV \\
    Higgs scalar mass  \(m_{5_2}\)         & 113.7~GeV
        & 0.75~GeV & 0.65~GeV \\
    \hline
    SO(10) unification scale  \(\Lambda_{\mathcal{O}}\) &
        $5.0 \cdot 10^{17}$~GeV & $0.55 \cdot 10^{17}$~GeV &
    $0.35 \cdot 10^{17}$~GeV \\
    D-Term $\sqrt{D_\mathcal{O}}$          & 30.0~GeV
        & \emph{fixed}  & \emph{fixed} \\
    matter scalar mass \(M_0\)             & 100~GeV
        & 2.2~GeV & 1.4~GeV \\
    \hline
    heaviest R-neutrino mass $M_{\nu_{R 3}}$ 
        & $6.30 \cdot 10^{14}$~GeV  &  $[0.86,50]\cdot 10^{14}$~GeV &
    $[1.8,24]\cdot 10^{14}$~GeV \\
    lightest neutrino mass \(m_{\nu_1}\)     &$3.5 \cdot 10^{-3}$~eV   &
          $[0.26,82]\cdot 10^{-3}$~eV & $[0.58,31]\cdot 10^{-3}$~eV\\
    \hline
  \end{tabular}
  \caption{\it High-scale SO(10) and SU(5) parameters in 2-step SO(10) $\to$ SU(5) $\to$
               SM breaking, and masses
            of the heavy 3-generation R-neutrino and of the lightest neutrino,
       as reconstructed from measurements at LHC and ILC assuming a fixed
       value for the D-term
   [The errors quoted correspond to 1$\sigma$].}
           \label{tab:ParameterDetermination2}
\end{table}
\begin{table}[!b]
  \centering
  \begin{tabular}{|l|c|c|c|}
    \hline
    Parameter    & Mass/Scale
        & Error for 500 fb$^{-1}$ &  Error for 1000 fb$^{-1}$\\
    \hline\hline
    SU(5) unification scale  \(\Lambda_{\mathcal U}\) & 2.35$\cdot 10^{16}$~GeV
        & 0.02$\cdot 10^{16}$~GeV & 0.02$\cdot 10^{16}$~GeV \\
    matter scalar mass \(m_{10,1}\)        & 163.3~GeV
        & 0.14~GeV  & 0.10~GeV \\
    matter scalar mass \(m_{\bar{5},1}\)   & 133.5~GeV
        & 0.55~GeV & 0.4~GeV \\
    matter scalar mass \(m_{10,3}\)        & 142.2~GeV
        & 0.85~GeV & 0.75~GeV \\
    matter scalar mass \(m_{\bar{5},3}\)   & 124.2~GeV
        & 0.65~GeV & 0.5~GeV \\
    Higgs scalar mass  \(m_{\bar{5}_1}\)   & 127.8~GeV
        & 0.22~GeV & 0.15~GeV \\
    Higgs scalar mass  \(m_{5_2}\)         & 113.7~GeV
        & 0.85~GeV & 0.7~GeV \\
    \hline
    SO(10) unification scale  \(\Lambda_{\mathcal{O}}\) &
        $5.0 \cdot 10^{17}$~GeV & \emph{fixed}  & \emph{fixed} \\
    D-Term $\sqrt{D_\mathcal{O}}$          & 30.0~GeV
        & 0.65 GeV & 0.5 GeV \\
    matter scalar mass \(M_0\)             & 100~GeV
        & 0.5~GeV & 0.4~GeV \\
    \hline
    heaviest R-neutrino mass $M_{\nu_{R 3}}$ 
        & $6.30 \cdot 10^{14}$~GeV  &  $[0.76,51]\cdot 10^{14}$~GeV &
    $[1.4,27]\cdot 10^{14}$~GeV \\
    lightest neutrino mass \(m_{\nu_1}\)     &$3.5 \cdot 10^{-3}$~eV   &
          $[0.25,90]\cdot 10^{-3}$~eV & $[0.51,44]\cdot 10^{-3}$~eV\\
    \hline
  \end{tabular}
  \caption{\it Same as Tab.~\ref{tab:ParameterDetermination2}, but for fixed
  SO(10) unification scale.}
           \label{tab:ParameterDetermination3}
\end{table}

In this way the complete set of couplings and mass parameters
could be analyzed. The results, with errors extrapolated from
Tab.~\ref{tab:experror2}, are shown in
Tab.~\ref{tab:ParameterDetermination2}. For illustration,
Tab.~\ref{tab:ParameterDetermination3} demonstrates how well the
parameters can be determined when the SO(10) unification scale
\(\Lambda_{\mathcal{O}}\) is identified with the string scale
while the SO(10) D-term is kept as a free variable. The differences
between the two sets of results are small.

The complex structure of the two-step breaking scenario, with
strong correlations between some parameters, makes it impossible
to derive {\it all} aspects of a general SO(10) scenario from
measurements without making any assumption on the SO(10) breaking
mechanism. Nevertheless, introducing assumptions about the SO(10)
breaking as exemplified in this section, the Terascale data allow
us to determine the two scales $\Lambda_{\mathcal U}$,
$\Lambda_{\mathcal{O}}$, the universal scalar mass $M_0$, and the
right-handed neutrino mass $M_{\nu_{R3}}$ of the third generation
independently. In particular, the two-step breaking scenario can
be clearly distinguished from the one-step scenario analyzed in
the previous section.

\section{Conclusions}\label{sec:conclusions}

If the roots of physics are located near the Planck scale,
experimental methods must be devised to explore the high-scale
physics scenario including the grand unification of the Standard
Model (SM) interactions up to, finally, gravity.

In this report we have studied two examples in which high-scale
parameters in supersymmetric SO(10) models have been connected
with experimental observations that could be expected in future
high-precision Terascale experiments at LHC and $e^+e^-$ linear
colliders.
For the sake of clarity, sectors that must be constructed
for a comprehensive SO(10) theory of all states but that are
not essential for the key points of our analysis, have not been
specified explicitly.
The renormalization group provides the tool for
bridging the gap between the Terascale experiments and the
underlying high-scale grand unification theory. Even though it
depends on the detailed values of the parameters
with which resolution the high-scale
picture can be reconstructed, an accurate picture could be
established in the example for one-step breaking SO(10) $\to$
SM, including the heavy mass of the right-handed
neutrino $\nu_{R3}$ expected in the seesaw mechanism.
As naturally anticipated, the analysis of
two-step breaking SO(10) $\to$ SU(5) $\to$ SM turns out to be
significantly more difficult, demanding a larger set of additional
assumptions before the parametric analysis can be performed.

The two examples have demonstrated nevertheless that
renormalization-group extrapolations based on high-precision 
results expected from Terascale experiments can provide essential
elements for the reconstruction of the physics scenario near the
grand unification if the theoretical frame of a comprehensive
SO(10) grand unified theory is specified.

\section*{Acknowledgements}
We thank P.~Bechtle,
S.~Wiesenfeldt and D.~Zerwas for discussions, and
A.~Djouadi for comments on the manuscript.
P.~M.~Zerwas thanks for the warm hospitality extended to him
at RWTH Aachen and Universit{\'e} Paris-Sud, Orsay.
This work is partially supported
by the German Ministry of Education and Research (BMBF) under
contract 05HT6WWA.
and by the U.S. Department of Energy, Division of
High Energy Physics, under Contract DE-AC02-06CH11357.


\begin{thebibliography}{99}

\bibitem{neutrino:oscillations}
Super-Kamiokande collaboration, Y.~Fukuda {\em et~al.},
  Phys. Rev. Lett. {\bf 81}, 1562 (1998), [hep-ex/9807003];
SNO collaboration, Q.~R. Ahmad {\em et~al.},
  Phys. Rev. Lett. {\bf 89}, 011301 (2002), [nucl-ex/0204008];
KamLAND collaboration, K.~Eguchi {\em et~al.},
  Phys. Rev. Lett. {\bf 90}, 021802 (2003), [hep-ex/0212021].

\bibitem{so10}
H.~Georgi,
  AIP Conf.\ Proc.\  {\bf 23} (1975) 575;
H.~Fritzsch and P.~Minkowski,
  Nucl.\ Phys.\  B {\bf 103} (1976) 61;


\bibitem{seesaw}
P.~Minkowski,
  Phys.\ Lett.\  B {\bf 67} (1977) 421;
T.~Yanagida,
{\it Proceedings, Workshop on the Baryon Number of the Universe
     and Unified Theories (Tsukuba 1979)};
M.~Gell-Mann, P.~Ramond and R.~Slansky,
  PRINT-80-0576-CERN;
R.~N. Mohapatra and G.~Senjanovic,
  Phys. Rev. Lett. {\bf 44}, 912 (1980);
J.~Schechter and J.~W.~F. Valle,
  Phys. Rev. {\bf D22}, 2227 (1980).

\bibitem{FPZ}
A.~Freitas, W.~Porod and P.~M.~Zerwas,
  Phys.\ Rev.\  D {\bf 72} (2005) 115002
  [arXiv:hep-ph/0509056].

\bibitem{Dterms}
H.~Baer, M.~A.~Diaz, P.~Quintana and X.~Tata,
  JHEP {\bf 0004} (2000) 016
  [arXiv:hep-ph/0002245].

\bibitem{A1}
  C.~S.~Aulakh,
  arXiv:0710.3945 [hep-ph].

\bibitem{G1}
  W.~Grimus and H.~Kuhbock,
  Eur.\ Phys.\ J.\  C {\bf 51}, 721 (2007)
  [arXiv:hep-ph/0612132].

\bibitem{A2}
  C.~S.~Aulakh,
  arXiv:hep-ph/0607252.

\bibitem{Aulakh:2006hs}
  C.~S.~Aulakh and S.~K.~Garg,
  arXiv:hep-ph/0612021.

\bibitem{Wies}
  S.~Wiesenfeldt, Phys.\ Rev.\ D{\bf 71} (2005) 075006
  [arXiv:hep-ph/0501223];
  S.~Wiesenfeldt and S.~Willenbrock,
  arXiv:0707.3300 [hep-ph].

\bibitem{Nath}
  P.~Nath and P.~Fileviez Perez,
  Phys.\ Rept.\  {\bf 441}, 191 (2007)
  [arXiv:hep-ph/0601023].

\bibitem{Babu}
  K.~S.~Babu and R.~N.~Mohapatra,
  Phys.\ Rev.\ Lett.\  {\bf 70}, 2845 (1993)
  [arXiv:hep-ph/9209215].

\bibitem{DreesDT}
M.~Drees,  Phys.\ Lett.\  B {\bf 181} (1986) 279.

\bibitem{Kolda:1996iw}
   C.~Kolda and S.P.~Martin, Phys.\ Rev.\ D{\bf 53} (1996) 3871.

\bibitem{JJK}
I.~Jack, D.~R.~T.~Jones and A.~F.~Kord,
  Phys.\ Lett.\  B {\bf 579} (2004) 180
  [arXiv:hep-ph/0308231].

\bibitem{spa}
J.~A.~Aguilar-Saavedra {\it et al.},
  Eur.\ Phys.\ J.\  C {\bf 46} (2006) 43
  [arXiv:hep-ph/0511344].

\bibitem{Pierce:1996zz}
D.~M.~Pierce, J.~A.~Bagger, K.~T.~Matchev and R.~J.~Zhang,
Nucl.\ Phys.\ B {\bf 491} (1997) 3
[arXiv:hep-ph/9606211].

\bibitem{MartinSP}
  S.~Heinemeyer, W.~Hollik and G.~Weiglein,
  Eur.\ Phys.\ J.\  C {\bf 9} (1999) 343
  [arXiv:hep-ph/9812472];
G.~Degrassi, P.~Slavich and F.~Zwirner,
Nucl.\ Phys.\ B {\bf 611} (2001) 403;
A.~Brignole, G.~Degrassi, P.~Slavich and F.~Zwirner,
Nucl.\ Phys.\ B {\bf 631} (2002) 195;
S.~P.~Martin,
  Phys.\ Rev.\  D {\bf 71} (2005) 116004
  [arXiv:hep-ph/0502168];
  Phys.\ Rev.\  D {\bf 72} (2005) 096008
  [arXiv:hep-ph/0509115].


\bibitem{GUTreconstruction}
 G.~A.~Blair, W.~Porod and P.~M.~Zerwas,
  Phys.\ Rev.\  D {\bf 63} (2001) 017703
  [arXiv:hep-ph/0007107];

\bibitem{GUTreconstruction2}
 G.~A.~Blair, W.~Porod and P.~M.~Zerwas,
   Eur.\ Phys.\ J.\  C {\bf 27} (2003) 263
  [arXiv:hep-ph/0210058].

\bibitem{Baer:2000hx}
  H.~Baer, C.~Balazs, J.~K.~Mizukoshi and X.~Tata,
  Phys.\ Rev.\  D {\bf 63} (2001) 055011
  [arXiv:hep-ph/0010068].

\bibitem{Dienes:1996du}
  K.~R.~Dienes,
  Phys.\ Rept.\  {\bf 287} (1997) 447
  [arXiv:hep-th/9602045].

\bibitem{Raby}
 S.~Raby, talk at the 15th International Conference on Supersymmetry and the
  Unification of Fundamental Interactions (SUSY07) in Karlruhe;
  arXiv:0710.2891 [hep-ph].

\bibitem{surv}
  C.~S.~Aulakh, B.~Bajc, A.~Melfo, A.~Rasin and G.~Senjanovic,
  Phys.\ Lett.\  B {\bf 460}, 325 (1999)
  [arXiv:hep-ph/9904352].

\bibitem{Cheng:1994bi}
  H.~C.~Cheng and L.~J.~Hall,
  Phys.\ Rev.\  D {\bf 51} (1995) 5289
  [arXiv:hep-ph/9411276].

\bibitem{Smir}
  E.~K.~Akhmedov, M.~Frigerio and A.~Y.~Smirnov,
  JHEP {\bf 0309}, 021 (2003)
  [arXiv:hep-ph/0305322].

\bibitem{Leptogen}
  S.~Davidson and A.~Ibarra,
    Phys.\ Lett.\  B {\bf 535} (2002) 25
    [arXiv:hep-ph/0202239];
    W.~Buchm\"uller, P.~Di Bari and M.~Pl\"umacher,
    Phys.\ Lett.\  B {\bf 547} (2002) 128
    [arXiv:hep-ph/0209301].

\bibitem{Aulakh:2007jm}
  C.~S.~Aulakh and S.~K.~Garg,
  arXiv:0710.4018 [hep-ph].

\bibitem{H}
  J.~Hisano and D.~Nomura,
  Phys.\ Rev.\  D {\bf 59} (1999) 116005
  [arXiv:hep-ph/9810479].

\bibitem{SPS}
B.~C.~Allanach {\it et al.},
  in {\it Proceedings, APS/DPF/DPB Summer Study on the Future of Particle Physics (Snowmass 2001),}
  and Eur.\ Phys.\ J.\ C {\bf 25} (2002) 113 [arXiv:hep-ph/0202233].

\bibitem{a}
   A.~Freitas, A.~von Manteuffel and P.~M.~Zerwas,
   Eur.\ Phys.\ J.\ C {\bf 34} (2004) 487
   [arXiv:hep-ph/0310182];
   A.~Freitas, H.~U.~Martyn, U.~Nauenberg and P.~M.~Zerwas,
   in {\it Proc. of the International Conference on Linear Colliders (LCWS 04),
   Paris, France, 19-24 Apr 2004}
   [arXiv:hep-ph/0409129].

\bibitem{c}
   K.~Desch, J.~Kalinowski, G.~Moortgat-Pick, M.~M.~Nojiri and G.~Polesello,
   JHEP {\bf 0402}, 035 (2004)
   [arXiv:hep-ph/0312069].

\bibitem{c1}
  S.~Y.~Choi, A.~Djouadi, M.~Guchait, J.~Kalinowski, H.~S.~Song and P.~M.~Zerwas,
  Eur.\ Phys.\ J.\  C {\bf 14} (2000) 535
  [arXiv:hep-ph/0002033];
  S.~Y.~Choi, A.~Djouadi, H.~S.~Song and P.~M.~Zerwas,
  Eur.\ Phys.\ J.\  C {\bf 8}, 669 (1999)
  [arXiv:hep-ph/9812236];
  S.~Y.~Choi, J.~Kalinowski, G.~A.~Moortgat-Pick and P.~M.~Zerwas,
  Eur.\ Phys.\ J.\  C {\bf 22} (2001) 563
  [Addendum-ibid.\  C {\bf 23} (2002) 769]
  [arXiv:hep-ph/0108117].

\bibitem{d}
   G.~Weiglein {\it et al.}  [LHC/LC Study Group],
   Phys.\ Rept.\  {\bf 426} (2006) 47
   [arXiv:hep-ph/0410364].

\bibitem{Bechtle:2004pc}
  P.~Bechtle, K.~Desch and P.~Wienemann,
  Comput.\ Phys.\ Commun.\  {\bf 174} (2006) 47
  [arXiv:hep-ph/0412012];
  P.~Bechtle, K.~Desch, W.~Porod and P.~Wienemann,
  Eur.\ Phys.\ J.\  C {\bf 46} (2006) 533
  [arXiv:hep-ph/0511006];
  R.~Lafaye, T.~Plehn and D.~Zerwas,
  arXiv:hep-ph/0512028 and
  arXiv:hep-ph/0404282;
R.~Lafaye, T.~Plehn, M.~Rauch and D.~Zerwas,
  arXiv:0709.3985 [hep-ph].

\bibitem{HA}
  K.~Desch, T.~Klimkovich, T.~Kuhl and A.~Raspereza,
  arXiv:hep-ph/0406229.

\bibitem{Bartl:1997yi}
  A.~Bartl, H.~Eberl, S.~Kraml, W.~Majerotto, W.~Porod and A.~Sopczak,
  Z.\ Phys.\  C {\bf 76} (1997) 549
  [arXiv:hep-ph/9701336];
 E.~Boos, H.~U.~Martyn, G.~A.~Moortgat-Pick, M.~Sachwitz, A.~Sherstnev and P.~M.~Zerwas,
  Eur.\ Phys.\ J.\  C {\bf 30} (2003) 395
  [arXiv:hep-ph/0303110].

\bibitem{CMZ}
 S.~Y.~Choi, H.~U.~Martyn and P.~M.~Zerwas,
  Eur.\ Phys.\ J.\  C {\bf 44} (2005) 175
  [arXiv:hep-ph/0508021].

\bibitem{Heinemeyer:2003ud}
  S.~Heinemeyer, W.~Hollik and G.~Weiglein, Phys. Rept. {\bf 425} (2006) 265
  [arXiv:hep-ph/0412214];
  B.C.~Allanach, A.~Djouadi, J.L.~Kneur, W.~Porod and P.~Slavich,
  JHEP {\bf 0409} (2004) 044 [arXiv:hep-ph/0406166];
  A.~Djouadi,
  arXiv:hep-ph/0503173;
  S.~Heinemeyer, S.~Kraml, W.~Porod and G.~Weiglein,
  JHEP {\bf 0309} (2003) 075
  [arXiv:hep-ph/0306181].

\bibitem{Spheno}
  W.~Porod,
  Comput.\ Phys.\ Commun.\  {\bf 153}, 275 (2003)
  [arXiv:hep-ph/0301101].

\bibitem{YM}
L.~Calibbi, Y.~Mambrini and S.~K.~Vempati,
  arXiv:0704.3518 [hep-ph].

\end{thebibliography}
\end{document}